%% file: ijcai26.tex
\documentclass{article}
\usepackage{ijcai26}

\usepackage{times}
\usepackage{soul}
\usepackage{url}
\usepackage[hidelinks]{hyperref}
\usepackage[utf8]{inputenc}
\usepackage[small]{caption}
\usepackage{graphicx}
\usepackage{amsmath}
\usepackage{amsthm}
\usepackage{booktabs}
\usepackage{algorithm}
\usepackage{algorithmic}
\usepackage[switch]{lineno}
\usepackage{multirow}
\usepackage{xcolor} 

\usepackage[table,xcdraw]{xcolor}

\title{
From Motion Signals to Insights: A Unified Framework for Student Behavior Analysis and Feedback in Physical Education Classes
}

\author{
Xian Gao$^1$
\and
Jiacheng Ruan$^1$\and
Jingsheng Gao$^1$\and
Mingye Xie$^1$\and
Zongyun Zhang$^1$\and \\
Ting Liu$^1$\And
Yuzhuo Fu$^1$\\
\affiliations
$^1$Shanghai Jiao Tong University\\
\emails
gaoxian@sjtu.edu.cn
}

\begin{document}

\maketitle

\input{src/1_abstract}
\input{src/2_introduction}

\input{src/3_related_work}
\input{src/4_methodology}
\input{src/5_experiments}
\input{src/6_case_study}
\input{src/7_conclusion}

\small
\bibliographystyle{named}
\bibliography{ijcai26.bib}

\input{src/appendix}

\end{document}

%% file: src/1_abstract.tex
\begin{abstract}

Analyzing student behavior in educational scenarios is crucial for enhancing teaching quality and student engagement. Existing AI-based models often rely on classroom video footage to identify and analyze student behavior. While these video-based methods can partially capture and analyze student actions, they struggle to accurately track each student's actions in physical education classes, which take place in outdoor, open spaces with diverse activities, and are challenging to generalize to the specialized technical movements involved in these settings. Furthermore, current methods typically lack the ability to integrate specialized pedagogical knowledge, limiting their ability to provide in-depth insights into student behavior and offer feedback for optimizing instructional design. To address these limitations, we propose a unified framework that integrates cascaded human activity recognition technologies based on motion signals, combined with advanced large language models (LLMs), to conduct more detailed analyses and provide feedback on student behavior in physical education classes. Starting from teachers’ instructional designs and students’ in-class IMU motion signals, the framework generates automated, pedagogically meaningful reports with targeted insights for optimizing learning and teaching. Experimental results demonstrate that our framework accurately identifies student behaviors, produces meaningful pedagogical insights, and leads to better student engagement in real-world PE classes.

\end{abstract}

%% file: src/2_introduction.tex
\begin{figure*}[!t]
    \centering
    \includegraphics[width=0.97\linewidth]{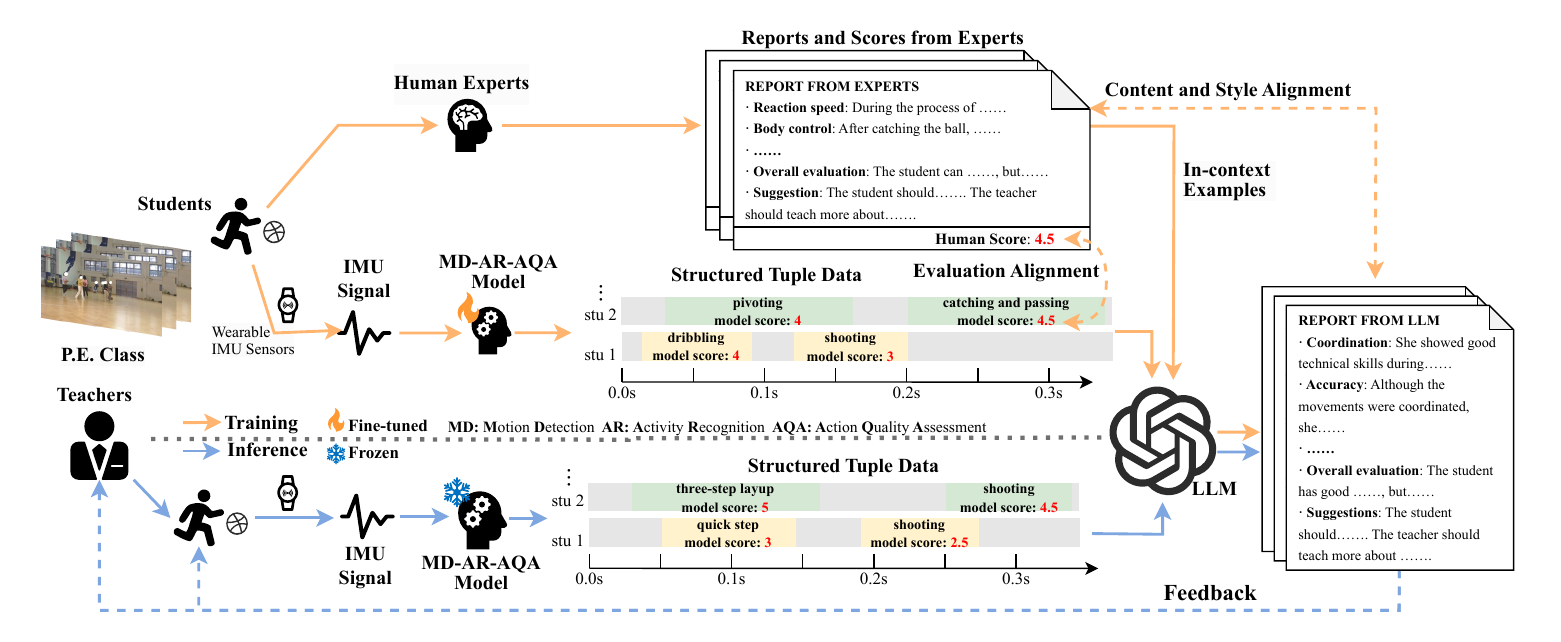}

    \caption{The overall architecture of our proposed framework. The cascaded motion detection, activity recognition, and action quality assessment models process students' IMU motion signals into structured tuple data (visualized as a timeline) readable by LLMs. The large language model then generates analytical reports as feedback for both students and teachers.}
    \label{fig:main}
\end{figure*}  

\section{Introduction}

Behavioral engagement is a key metric for evaluating student performance, reflecting actions such as task completion and participation \cite{fredricksSchoolEngagementPotential2004}. In physical education, its importance is heightened as it impacts physical development, skill acquisition, teamwork, and mental health  \cite{baileyPhysicalEducationSport2006,yuPhysicalActivitySelfefficacy2024}, while also indicating alignment with instructional goals. Traditional behavior analysis methods, reliant on observation or video review, are labor-intensive and insufficient for accurately capturing each student’s behavior in the dynamic context of physical education, leading to undetected disengagement and compromised learning experiences.

Existing research has utilized computer vision technology for automated analysis of student behavior in classroom videos, including action classification \cite{zhengIntelligentStudentBehavior2020,yangStudentClassroomBehavior2023}, pose estimation \cite{linStudentBehaviorRecognition2021,yuCCPoseNetHumanPose2023}, and emotion recognition \cite{liExploringArtificialIntelligence2023,huRealTimeClassroomBehavior2024}. However, applying these video-based methods to physical education scenarios faces two core challenges: (1) Outdoor open spaces, large-scale movements, and specialized technical actions make it difficult for video-based methods to comprehensively and accurately capture student behaviors. (2) Most current approaches focus solely on action recognition and classification, with limited ability to integrate educational knowledge and provide targeted suggestions for teaching optimization.

To address these issues, we utilize IMU motion signals, which are commonly used in sports research and are able to precisely track individual motion trajectories, instead of video data, to analyze student behavior in physical education classes. A unified framework is proposed, leveraging human activity recognition and LLMs to generate insightful teaching reports from raw motion signal segments. Specifically, the framework first detects active motion segments from raw signals, then classifies these segments and evaluates their quality using activity recognition techniques, converting non-textual signal data into LLM-readable textual statistical summaries. Finally, guided by education-specific prompts, LLMs generate comprehensive analytical reports for both individual students and the entire class. To improve generalization across diverse physical education activities, the framework is pre-trained on large-scale public and synthetic datasets, ensuring robust performance with minimal fine-tuning.

Our primary contributions are as follows: 
\begin{itemize}
\item We propose a novel approach for analyzing motion signals using large language models, which cannot directly process raw signal data. To the best of our knowledge, this is the first attempt to transform motion signals into text information readable by LLMs.

\item We introduce a unified framework that transforms raw physical education class motion signals into insightful teaching analytics. Our generalizable approach can accurately identify a wide range of student movements in physical education classes with only a small amount of sample data, providing comprehensive and in-depth analysis and feedback through integration with LLMs. 

\item We validate the effectiveness of our approach using basketball classes by a case study and real-class experiment, and both quantitative and qualitative experimental results confirm the robustness and efficacy of our framework.

\end{itemize}

%% file: src/3_related_work.tex
\section{Related Work}
\subsection{AI-based Student Behavior Analysis}

Current AI-based methods for student behavior analysis primarily utilize computer vision to examine classroom videos. Most studies treat behavior analysis as an object detection task (e.g., identifying hand-raising or sleeping \cite{zhouWhoAreRaising2018,liSleepGestureDetection2019,liuFastAccurateHandRaising2020}) or detect multiple behaviors \cite{wangEffectiveYawnBehavior2019,yangStudentClassroomBehavior2023} by integrating pose estimation, micro-expression, or emotion recognition \cite{peiMicroexpressionRecognitionAlgorithm2019,linStudentBehaviorRecognition2021,liExploringArtificialIntelligence2023,huRealTimeClassroomBehavior2024}. Temporal action detection has also been explored to capture the duration of student actions \cite{yuRawVideoPedagogical2024}. Unlike these video-based approaches, our method analyzes students’ IMU motion signals, making it more suitable for the open and complex movements typical of physical education classes.

\subsection{IMU-based Human Activity Recognition}
Human Activity Recognition (HAR) identifies and understands human behaviors using IMU sensors, which capture accelerometer and gyroscope data and are widely used in smartphones and wearable devices. Existing studies employ deep learning techniques (e.g., CNN \cite{zengConvolutionalNeuralNetworks2014,ronaoHumanActivityRecognition2016,huangShallowConvolutionalNeural2021}, RNN \cite{muradDeepRecurrentNeural2017,inoueDeepRecurrentNeural2018}, and LSTM \cite{guanEnsemblesDeepLSTM2017,ashryCHARMDeepContinuousHuman2020}) or large-scale pre-training with contrastive learning \cite{moonIMU2CLIPMultimodalContrastive2022,girdharImageBindOneEmbedding2023a,li-etal-2025-sensorllm} to process IMU signals. In this paper, we align IMU signals with textual modality (action type labels) to enhance generalization in physical education scenarios.

%% file: src/4_methodology.tex
\begin{figure*}[!t]
    \centering
    \includegraphics[width=0.95\linewidth]{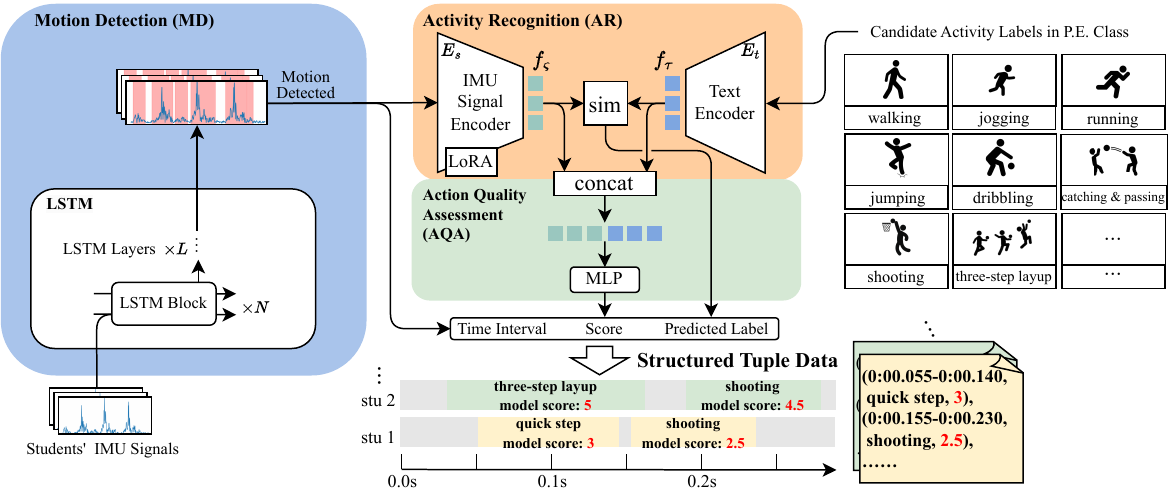}
    \caption{The architecture and corresponding processing workflows of MD, AR, and AQA model.}
    \label{fig:model}
\end{figure*}

\section{Methodology}

\subsection{Framework Overview}

Our framework begins with the instructional design and execution of a physical education class. Wearable IMU sensor devices (e.g., smartphones or wristbands) are employed to collect motion signals from each student throughout the class. Given that the collected signals contain both active movement segments (e.g., running, jumping, and other specialized actions) and static segments of inactivity, student behavior analysis should focus on the active segments. Therefore, motion detection is performed to segment and remove redundant parts of the sequence. Subsequently, the identified active segments undergo activity recognition to classify student behaviors, followed by action quality assessment to evaluate performance, which enables a deeper analysis of the students' technical performance.

In our framework, as illustrated by the \textcolor[RGB]{126,166,224}{blue workflow} in Figure \ref{fig:main}, these three steps are executed through a cascaded MD-AR-AQA model, comprising \textbf{M}otion \textbf{D}etection, \textbf{A}ctivity \textbf{R}ecognition, and \textbf{A}ction \textbf{Q}uality \textbf{A}ssessment. The signals are sequentially processed to generate structured (motion time interval, predicted action label, action score) tuples, converting non-textual signal data into text interpretable by large language models. These structured data, combined with predefined prompts, are input into a large language model to generate personalized, in-depth feedback and recommendations for both students and teachers.

\subsection{Training Process}

The training process is illustrated by the \textcolor[RGB]{255,181,112}{orange workflow} in Figure \ref{fig:main}. We conducted pre-training of the motion detection and activity recognition models on both synthetic and publicly available datasets. When applied in real-world physical education class scenarios, a small amount of annotated and scored example signals, involving professional movements from the class, is required to fine-tune the action recognition model and train the action quality assessment model. This step ensures that the models can accurately recognize specialized actions in specific scenarios and align the evaluation model with human assessments. After training the cascaded MD-AR-AQA model, the structured tuple outputs generated by the model are combined with human expert evaluation reports (typically provided by physical education teachers). These are used as in-context examples, alongside predefined prompts, to be input into the large language model. The prompts are iteratively refined to align the content and style of the model-generated report with that of the expert-written reports. Through this process, our framework is capable of autonomously generating student behavior analysis reports in specific physical education settings, closely mirroring human evaluations. The training details of each module will be elaborated in the following sections.

\section{Technical Architecture}

\subsection{Motion Detection Model}

\subsubsection{Model Overview}
The motion detection model is employed to identify and segment the active portions of the signal. This process can be framed as a sequence labeling task, where the model takes a motion sequence as input and produces a binary sequence indicating whether each time point corresponds to a motion event. We segment each sequence using a fixed window length 
$L_w$ and a sliding window step size $L_s$. An LSTM., as shown in Figure \ref{fig:model}, is then used to detect motion segments within each window, predicting the probability that each frame belongs to a motion signal. The final prediction for each frame is obtained by averaging the predictions across all sliding windows that include that frame.

\subsubsection{Training Details}
Due to the absence of publicly available datasets for IMU signal motion detection, we created a synthetic dataset by sampling and concatenating motion segments from activity recognition datasets. We first categorized actions into two types: motion (e.g., walking, running) and stationary (e.g., sitting, standing), and relabeled them with binary annotations. Random samples from these categories were concatenated to form IMU signal sequences encompassing different motion states, with corresponding ground truth labels. Linear interpolation was applied at transition points to ensure smooth shifts between actions. This process generated a labeled dataset of varying lengths and motion states, ensuring diversity in sequence length and content. The data was used to train the motion detection model, which outputs a binary sequence indicating whether each time point corresponds to a motion segment.

\subsection{Activity Recognition Model}

\subsubsection{Model Overview}

The activity recognition model classifies the detected motion segments. To improve its generalization for diverse student movements in physical education classes, we employ a pre-training and fine-tuning approach. In the pre-training phase, contrastive learning is used to align the IMU signal modality of a custom encoder with the text modality. By mapping the IMU signal encoder's feature space to the frozen feature space of the CLIP \cite{radfordLearningTransferableVisual2021a} text encoder, we fully utilize the textual knowledge embedded in CLIP. After pre-training, we fine-tune the model on a dataset of movements specific to physical education, resulting in an aligned model capable of activity recognition. As illustrated in Figure \ref{fig:model}, the model consists of an IMU signal encoder $E_s$ and a text encoder $E_t$, which encode the IMU signal input $\varsigma$ and text input $\tau$, producing feature vectors $f_{\varsigma} = E_{s}(\varsigma)$ and $f_{\tau} = E_{t}(\tau)$. As the encoders are modality-aligned, the similarity between an IMU signal and its corresponding text label is maximized, while the similarity with other labels is minimized. Therefore, for an IMU signal $\varsigma$, the predicted label $\hat{\tau}$ is determined by measuring its similarity with each candidate label in the action type set $\Lambda$. This can be expressed mathematically as:

{\small
\begin{equation}
\hat{\tau} = \mathop{\arg\max}_{\tau_{i} \in \Lambda} {\rm sim}(f_{\varsigma},f_{\tau_{i}})= \mathop{\arg\max}_{\tau_{i}\in \Lambda} \frac{E_{s}(\varsigma)\odot E_{t}(\tau_{i})}{\|E_{s}(\varsigma)\|\cdot\|E_{t}(\tau_{i})\|}
\end{equation}
}

\subsubsection{Pre-training Phase}

We employ contrastive learning for the pre-training of the activity recognition model. By collecting nearly all publicly available human action recognition datasets, we built a large-scale pre-training dataset to train the Transformer-based signal encoder $E_s$. Through contrastive learning, its feature vectors are mapped into the feature space of the pre-trained CLIP text encoder $E_t$ and aligned with the features of the corresponding labels.

\subsubsection{Fine-tuning Phase}

We apply the LoRA method \cite{huLoRALowRankAdaptation2021} to fine-tune the activity recognition model. This approach involves freezing the pre-trained model's weights and adding trainable LoRA blocks to the Transformer layers of the IMU signal encoder. The fine-tuning data consists of IMU signal segments from real physical education classes, annotated with action categories. This data can be obtained by demonstrating specific actions, recording the signals, and segmenting them using the pre-trained motion detection model. The signal segments and text labels are then used for LoRA fine-tuning, continuing the contrastive learning approach from pre-training. As the model has already been pre-trained on large datasets, minimal fine-tuning with a small number of real-world data is sufficient for high-accuracy activity recognition.

\subsection{Action Quality Assessment Model} 
\subsubsection{Model Overview}

Action quality assessment aims to score input activity signals, generating outputs that closely resemble human evaluations, thereby enriching the information for report generation. This process mimics the way human experts assess students' performance. The model concatenates the feature vector $f_{\varsigma}$ of the activity signal $\varsigma$, produced by the activity recognition model, with the feature vector $f_{\hat{\tau}}$ of the corresponding recognition result $\hat{\tau}$ as the input. A multilayer perceptron (MLP) is then employed to map these feature vectors to a numerical score between 0 and 5. Formally, this can be expressed as:

\begin{equation}
{\rm score} = {\rm MLP}({\rm concat}(f_{\varsigma}, f_{\hat{\tau}}))
\end{equation}

\subsubsection{Training Details}

The data for training the action quality assessment model is the same as that used for fine-tuning the activity recognition model, with annotations scoring the normative accuracy of the actions. After training the recognition model, the frozen signal and text encoders extract features from the training data. These feature vectors are concatenated and input into the quality assessment model, which learns to predict scores from the input. As a lightweight MLP-based model, it requires only a small amount of training data to effectively represent action quality.

\subsection{Large Language Model Integration}

In the final stage of the framework, a large language model is employed to generate reports based on the structured text output from the cascaded MD-AR-AQA models. Through motion detection, activity recognition, and action quality evaluation, we obtain the start and end times, action types, and quality scores for each student's actions during the class. These are represented as structured text-based (time interval, action type label, action score) tuples, which are fed into the large language model along with appropriate prompts to generate analytical reports. As model performance and prompt design directly influence the quality of the generated reports \cite{liuPretrainPromptPredict2023,liGuidingLargeLanguage2023,sahooSystematicSurveyPrompt2024}, we utilized the GPT-4o model along with carefully crafted, templated prompts to enhance the model's ability to follow instructions and analyze data. This approach ensures the generation of reports that are enriched with expert-level insights and closely resemble the style of human experts.

%% file: src/5_experiments.tex
\section{Experiments}
\subsection{Datasets and Metrics}

The synthetic training dataset for the motion detection model was generated based on the KU-HAR dataset \cite{sikderKUHAROpenDataset2021}. A dataset of 100,000 samples was synthesized following the data synthesis methods outlined in the Technical Architecture section. Evaluation metrics, including precision, recall, and F1-score, were employed to assess the model’s performance in segmenting and identifying intervals of activity sequences.

The pre-training dataset for the activity recognition model was collected from publicly available datasets, including KU-HAR \cite{sikderKUHAROpenDataset2021}, RealWorld2016 \cite{sztylerOnbodyLocalizationWearable2016}, and MotionSense \cite{malekzadehProtectingSensoryData2018,malekzadehPrivacyUtilityPreserving2019}. After data cleaning and augmentation, the final dataset contained approximately 0.75M samples. Notably, the UCI-HAR dataset \cite{anguitaPublicDomainDataset2013} was reserved for fine-tuning and zero-shot testing and thus excluded from the training set.

The fine-tuning of the activity recognition model and the training of the action quality assessment model were conducted from 320 real-world samples across 12 action categories. Evaluation metrics for the activity recognition model included classification accuracy and F1-score, while the action quality assessment model was evaluated using mean squared error (MSE) and Pearson correlation coefficient, compared to human assessments.

\subsection{Implementation Details}

The LSTM-based motion detection model comprises 6 LSTM layers with a hidden size of 64, and the sliding window step size $L_s$ is set to 1/4 of the window length $L_w$, ensuring each segment appears in 4 windows. The pre-trained activity recognition model's IMU signal encoder utilizes a 6-layer Transformer architecture. The action quality assessment model is structured as a 3-layer MLP. During fine-tuning, the LoRA rank is set to 32, with 16 samples per class for few-shot fine-tuning. All experiments were conducted on NVIDIA A100 80GB GPUs, and the large language model employed during inference was GPT-4o.

\subsection{Main Results}
\subsubsection{Motion Detection}

The average precision, recall, and F1-score of our motion detection model on the test set are 91.63\%, 92.41\%, and 91.82\%, respectively. These high performance metrics indicate that our model accurately segments motion sequences from the raw signals, with a low probability of misclassifying non-motion segments as containing motion.

\subsubsection{Activity Recognition}

\begin{table}[!t]
\centering
\small
\begin{tabular}{c|cc|cc}
\hline
\multirow{2}{*}{Category} & \multicolumn{2}{c|}{ Zero-shot} & \multicolumn{2}{c}{ Fine-tune} \\
        & Acc. \% & F1 \% & Acc. \% & F1 \% \\ \hline
Overall & 54.09    & 47.95    & 92.30    & 92.31    \\ \hline
Walking    & 66.53    & 68.46    & 97.98    & 97.10    \\
Walking Upstairs      & 65.18    & 53.25    & 91.93    & 93.93    \\
Walking Downstairs    & 32.38    & 42.50    & 93.57    & 92.25    \\
Sitting     & 0.00        & 0.00        & 85.74    & 87.44    \\
Standing   & 89.29    & 64.89    & 91.92    & 90.81    \\
Laying     & 64.43    & 58.59    & 92.74    & 92.31    \\ \hline
\end{tabular}

    \caption{Activity recognition accuracy and F1-score on UCI-HAR dataset.}
    \label{UCI-Metrics}
\end{table}

\begin{figure}[!t]
    \centering
    \includegraphics[width=\linewidth]{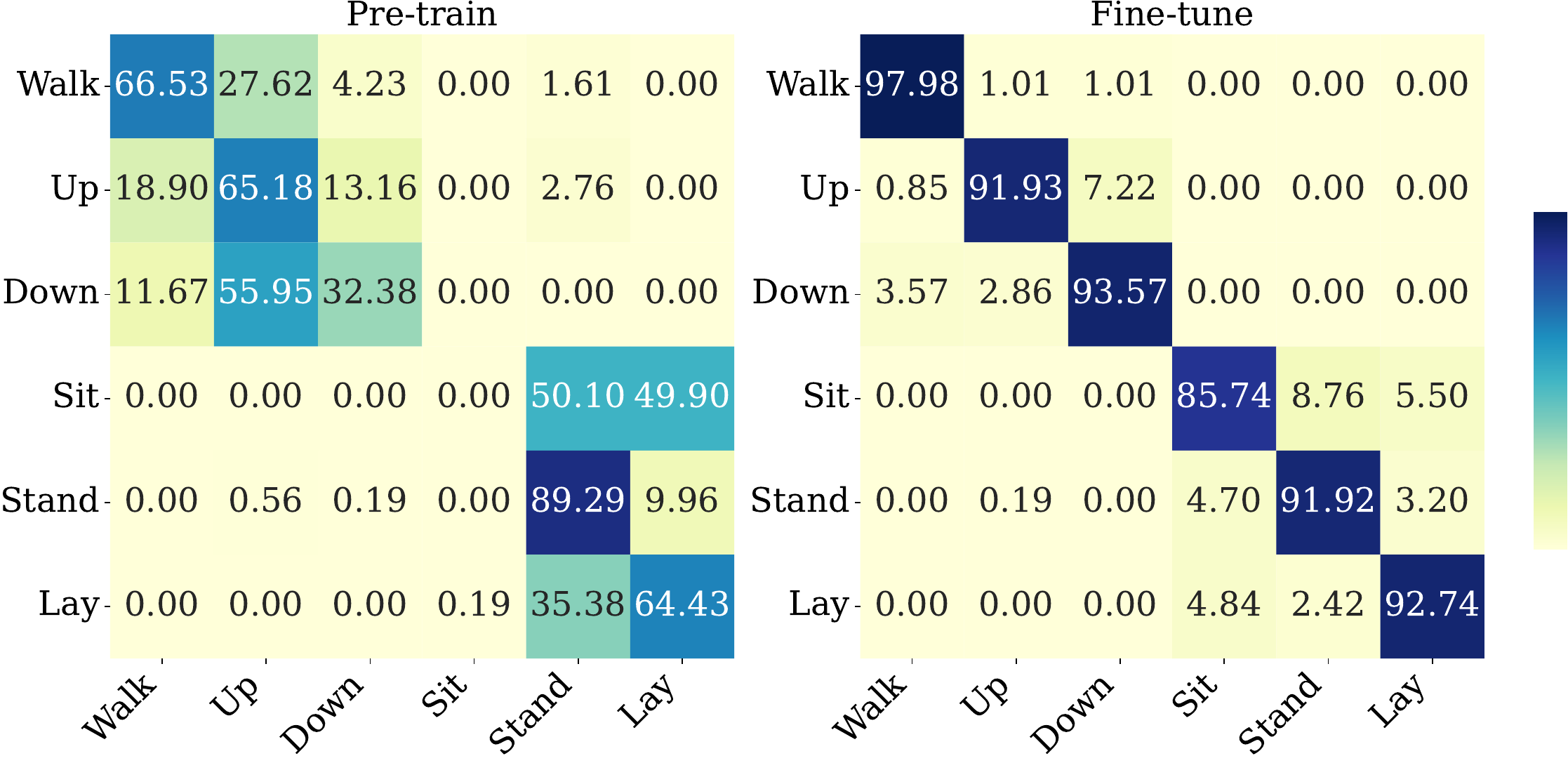}
    \caption{Heatmap of the confusion matrices.}
    \label{fig:uci-heatmap}
\end{figure}

Table \ref{UCI-Metrics} presents the zero-shot and fine-tuned accuracy and F1-score of our model on the UCI-HAR test set. Figure \ref{fig:uci-heatmap} visualizes the confusion matrix as a heatmap, illustrating classification performance before and after fine-tuning. After pre-training, the model achieves a zero-shot classification accuracy of 54.09\% on the UCI-HAR dataset. The model initially struggles to distinguish between similar actions, such as "Walking", "Walking Upstairs", and "Walking Downstairs" and between "Sitting", "Standing", and "Laying" due to the similarity in their signals. However, fine-tuning significantly improves classification, yielding an overall accuracy of 92.30\% and an F1-score of 92.31\%. Table \ref{UCI-SOTA-Metrics} compares our model with state-of-the-art methods, demonstrating its superior performance.

\begin{table}[!t]
\centering
\small
\begin{tabular}{ccc}
\hline
\multicolumn{1}{c|}{Method}             &  Acc. \%     &  F1 \%   \\ \hline
\multicolumn{3}{c}{Zero-shot}                                               \\ \hline
\multicolumn{1}{c|}{IMU2CLIP \cite{moonIMU2CLIPMultimodalContrastive2022}}           &         29.66         &       15.49         \\
\multicolumn{1}{c|}{ImageBind \cite{girdharImageBindOneEmbedding2023a}}          &        16.76          & 10.26     \\
\multicolumn{1}{c|}{\textbf{ours}}      & \textbf{54.09}    & \textbf{47.95} \\ \hline
\multicolumn{3}{c}{Fine-tune}                                               \\ \hline
\multicolumn{1}{c|}{Barlow \cite{zbontarBarlowTwinsSelfSupervised2021a}}             & \textbackslash{} & 57.50           \\
\multicolumn{1}{c|}{DCL \cite{chuangDebiasedContrastiveLearning2020}}                & \textbackslash{} & 89.80           \\
\multicolumn{1}{c|}{HardDCL \cite{robinsonContrastiveLearningHard2021}}            & \textbackslash{} & 91.00             \\
\multicolumn{1}{c|}{COCOA \cite{deldariCOCOACrossModality2022}}              & \textbackslash{} & 90.90          \\
\multicolumn{1}{c|}{NCE \cite{oordRepresentationLearningContrastive2019}}                & \textbackslash{} & 91.40           \\
\multicolumn{1}{c|}{TS2ACT (w.o. image) \cite{xiaTS2ACTFewShotHuman2023a}} & 86.30            & 85.90           \\
\multicolumn{1}{c|}{\textbf{ours}}      & \textbf{92.30}   & \textbf{92.31} \\ \hline
\end{tabular}

    \caption{Compared with other IMU-text cross-modal transfer learning methods.}
    \label{UCI-SOTA-Metrics}
\end{table}

Table \ref{AR-Class-Metrics} shows the performance of our model on real-world physical education class data. Since the pre-training data lacks specialized technical movements typical of physical education, the model performs well in zero-shot recognition for general categories present in the training set (e.g., Jumping, Running, Jogging) but shows lower accuracy for specialized movements (e.g., Three-step Layup, Starting and Stopping). However, the model exhibits some generalization to technical actions like pivoting and dribbling. After fine-tuning with few-shot learning, it effectively captures the characteristics of these specialized movements, significantly enhancing recognition accuracy and validating the representational strength acquired during pre-training.

\begin{table}[!t]
\centering
\small
\begin{tabular}{c|cc|cc}
\hline
\multirow{2}{*}{Category} & \multicolumn{2}{c|}{Zero-shot}      & \multicolumn{2}{c}{Few-shot} \\
                          & Acc. \%                       & F1 \%    & Acc. \%         & F1 \%         \\ \hline
Overall                   & \multicolumn{1}{c}{23.44}  & 17.54 & 79.69         & 80.24        \\ \hline
Jumping                   & \multicolumn{1}{c}{100.00} & 53.33 & 100.00        & 100.00       \\
Running                   & \multicolumn{1}{c}{60.00}  & 50.00    & 60.00         & 75.00        \\
Jogging                   & \multicolumn{1}{c}{50.00}  & 37.50  & 83.33         & 71.43        \\
Walking                   & \multicolumn{1}{c}{40.00}  & 20.00    & 60.00         & 66.67        \\
Pivoting                  & \multicolumn{1}{c}{40.00}  & 36.36 & 60.00         & 75.00        \\
Dribbling                 & \multicolumn{1}{c}{20.00}  & 13.33 & 100.00        & 100.00       \\
Catching \& Passing       & \multicolumn{1}{c}{0.00}      & 0.00     & 66.67         & 72.73        \\
Quick-step                & \multicolumn{1}{c}{0.00}      & 0.00     & 100.00        & 94.12        \\
Shooting                  & \multicolumn{1}{c}{0.00}      & 0.00     & 75.00         & 85.71        \\
Sliding                   & \multicolumn{1}{c}{0.00}      & 0.00     & 85.71         & 75.00        \\
Starting \& Stopping      & \multicolumn{1}{c}{0.00}      & 0.00     & 80.00         & 61.54        \\
Three-step Layup          & \multicolumn{1}{c}{0.00}      & 0.00     & 75.00         & 85.71        \\ \hline
\end{tabular}
    \caption{Activity recognition accuracy of each label category on real motion signals from physical education class.}
    \label{AR-Class-Metrics}
\end{table}

\subsubsection{Action Quality Assessment}


Table \ref{QA-Metrics} presents a comparison between our quality assessment model and the state-of-the-art video-based human behavior understanding model, MotionLLM \cite{chenMotionLLMUnderstandingHuman2024}, in the task of action quality assessment. The results indicate that our model achieves a lower mean squared error (MSE) compared to human ratings and demonstrates a higher correlation, suggesting that the scoring of our quality assessment model aligns more closely with human preferences.

\begin{table}[!t]
\centering
\small
    \begin{tabular}{c|c|c}
    \hline
   Model         & MSE $\downarrow$        & Pearson Corr. $\uparrow$  \\ \hline 
   Human          & 0.00           & 1.00           \\
    MotionLLM \cite{chenMotionLLMUnderstandingHuman2024}        & 1.10          & 0.08           \\
    \textbf{ours}     & \textbf{0.63} & \textbf{0.51} \\ \hline
    \end{tabular}
    \caption{Performance of the quality assessment model compared with video-based SOTA method.}
    \label{QA-Metrics}
\end{table}

\begin{table}[!t]
\centering
\small
\begin{tabular}{c|ccc}
\hline
$L_s$ / $L_w$   & Avg. Precision \% & Avg. Recall \%   & Avg. F1 \%       \\ \hline
1             & 89.13          & 90.31          & 89.96          \\
0.5           & 89.77          & 91.27          & 90.03          \\
0.25 & 91.63 & 92.41 & 92.02 \\ \hline
\end{tabular}
    \caption{The average metrics on the test set under different sliding window lengths. 
    }

    \label{MD-step-Metrics}
\end{table}

\begin{table}[!t]
\centering
\small
\begin{tabular}{c|c|cccc}
\hline
\multirow{2}{*}{\begin{tabular}[c]{@{}c@{}}LoRA \\ Rank\end{tabular}} &
  \multirow{2}{*}{\begin{tabular}[c]{@{}c@{}}Pre-trained \\ Model\end{tabular}} &
  \multicolumn{4}{c}{\# K-shot} \\ \cline{3-6} 
   &                        & 2     & 4     & 8     & 16    \\ \hline
4  & \multirow{4}{*}{23.44} & 29.69 & 45.31 & 53.12 & 68.75 \\
8  &                        & 35.94 & 53.12 & 57.81 & 71.88 \\
16 &                        & 45.31 & 56.25 & 65.62 & 73.44 \\
32 &                        & 43.75 & 54.69 & 67.19 & 79.69 \\ \hline
\end{tabular}
    \caption{Activity recognition accuracy on different k-shot and LoRA rank settings.}
    \label{AR-Metrics}
\end{table}

\subsection{Ablation Studies}
\subsubsection{Sliding Window Step Size in Motion Detection}

Table \ref{MD-step-Metrics} presents the average performance metrics of the motion detection model on the test set with varying sliding window step sizes $L_s$. A smaller step size increases the overlap between windows, enabling the averaging of predictions across multiple windows, which enhances motion detection accuracy. However, this also raises the number of windows and thus the computational cost. To balance efficiency and performance, we choose a sliding window step size of $L_s = 0.25 * L_w$ in this study.

\subsubsection{Scale of K-shot and LoRA Rank}

Table \ref{AR-Metrics} presents the performance of few-shot fine-tuning across different LoRA ranks and shot numbers. While higher LoRA ranks combined with small sample sizes can lead to overfitting, overall performance improves with increased sample size and model scale, confirming the effectiveness of our LoRA fine-tuning method. In practical use, additional fine-tuning data is expected to further boost model performance.

%% file: src/6_case_study.tex
\begin{figure}[!t]
    \centering
    \includegraphics[width=0.95\linewidth]{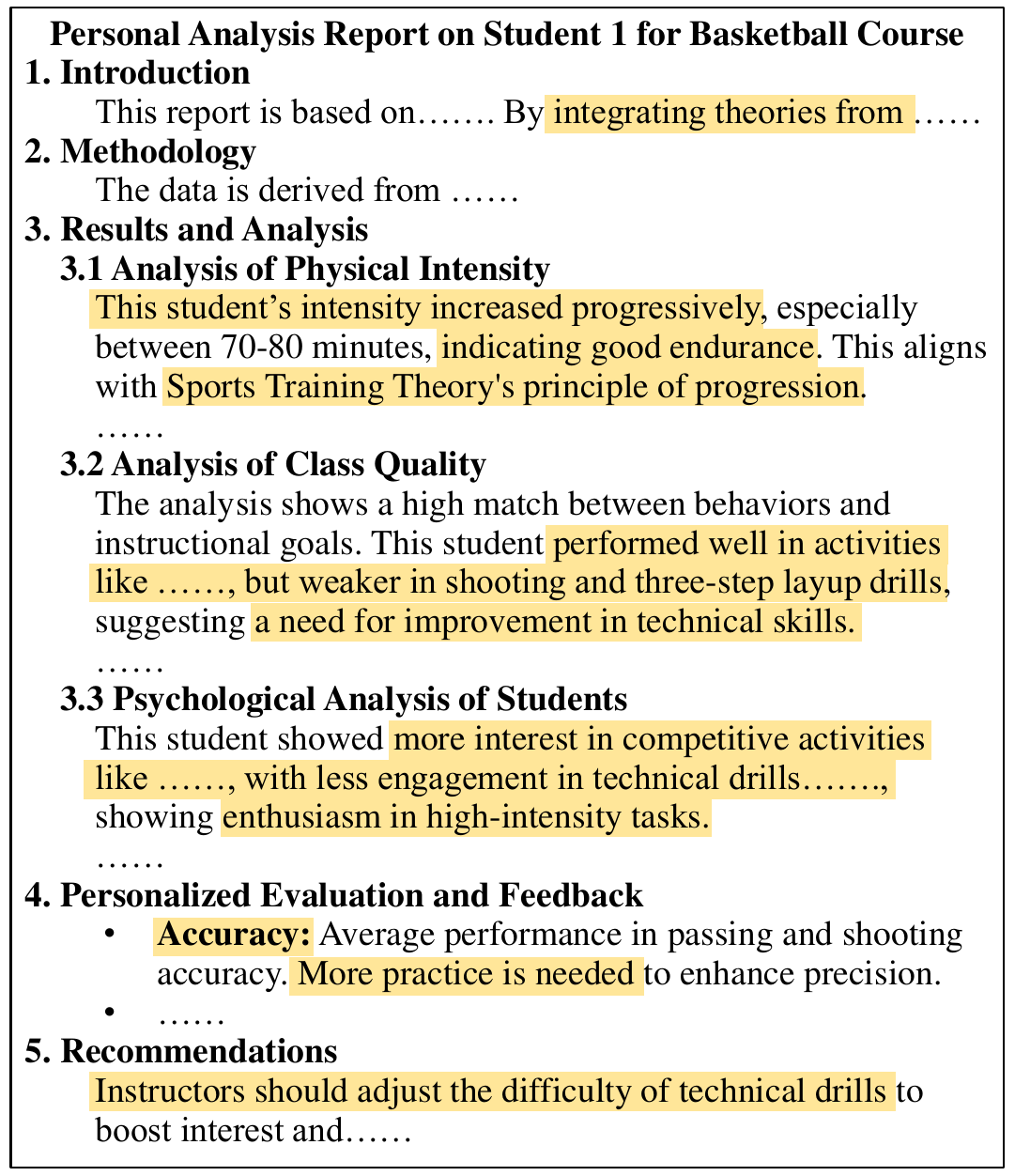}
    \caption{An excerpt from a personal analysis report generated by LLM.  The highlighted sections indicate content that human evaluators identified as pedagogically significant.}
    \label{fig:report-individual}
\end{figure}

\begin{figure}[!t]
    \centering
    \includegraphics[width=0.95\linewidth]{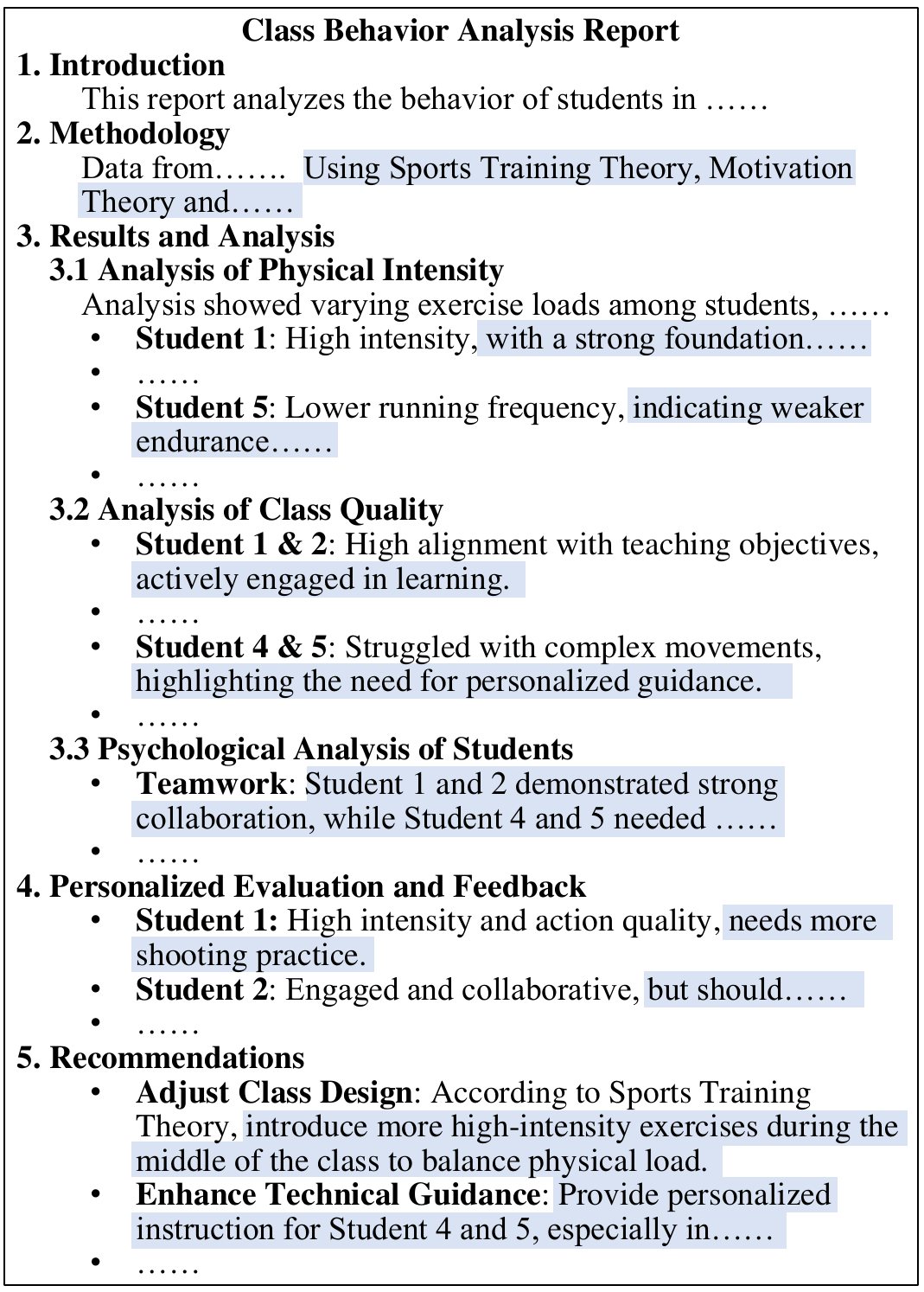}
    \caption{An excerpt from a class analysis report generated by LLM. 
    }
    \label{fig:report-class}
\end{figure}

\section{Case Study}

We collected students' motion signals and teachers' instructional designs from real physical education classes and applied our framework for comprehensive analysis, generating behavior reports for both individuals and the class. These reports were evaluated by instructors and physical education experts. Through iterative refinement of prompts and report formats, we produced pedagogically meaningful content. All evaluators, including professors, instructors, and teaching assistants, held PhDs in sports education or related fields. Each student's action scoring and report analysis was conducted by no fewer than two experts. The evaluation of the reports was based on the consensus of the evaluators. In cases where there were disagreements among evaluators, we invited an additional evaluator, who either participated in or closely watched the video of the class, to make the final judgment.

\subsection{Analysis Report for Student and Class}

Figure \ref{fig:report-individual} shows an excerpt from an LLM-generated individual student analysis report that evaluates behavior across three key dimensions—physical exertion, class quality, and psychological factors such as engagement and cooperation—and provides personalized feedback and teaching suggestions. he highlighted sections indicate content that the evaluators deemed pedagogically significant, such as aspects related to physical education and assessment, analyses grounded in sound teaching theory, and feedback or recommendations that could potentially improve instructional design and enhance the student learning experience.

Figure \ref{fig:report-class} shows an excerpt from the LLM-generated comprehensive class analysis report, which aggregates individual reports and adds interpersonal analyses, such as classifying students by participation level and examining team dynamics during group activities. Based on integrated class-level data, the model offers instructional design recommendations (e.g., adding higher-intensity activities mid-class) and practical strategies to enhance engagement and learning, demonstrating the framework’s ability to convert raw motion signals into actionable insights and solutions.

In the appendix of our supplementary materials, we provide additional comprehensive reports along with expert evaluations, as well as explanations regarding the alignment between the report content and the instructional plan.

\subsection{Real Class Experiment}

To validate the framework’s practical effectiveness, a real classroom experiment was conducted at a QS Top 50 university in China with 72 student volunteers (37 freshmen, 35 sophomores) from basketball PE classes, focusing on moving catching/passing and the three-step layup. The experiment followed the proposed cycle: IMU sensors collected motion data in the first two sessions, the framework generated analytical reports, and teachers adjusted student grouping for the subsequent two sessions, using consistent data collection before and after optimization. Student engagement was quantitatively evaluated by comparing the mean proportional frequency and duration, along with the coefficients of variation, of key instructional actions.

\begin{table}[!t]
\centering
\small
\begin{tabular}{c|c|c}
\hline
Indicators & Before (\%) & After (\%)      \\ \hline
Core Action Prop. $\uparrow$                  & 53.28         & 57.68 (+4.40) \\
Core Action Duration Prop.   $\uparrow$         & 67.01       & 72.47 (+5.46)   \\
CV of Core Actions $\downarrow$   & 78.42         & 56.17 (-22.25)  \\
CV of Active Duration     $\downarrow$      & 65.29         & 48.73 (-16.56)  \\ \hline
\end{tabular}%
\caption{Comparison of Key Indicators Before and After Feedback and Teaching Optimization.}
\label{tab:real}
\end{table}

\begin{table}[!t]
\centering
\small
\begin{tabular}{c|cc|cc}
\hline
\cellcolor[HTML]{FFFFFF}{\color[HTML]{333333} } &
  \multicolumn{2}{c|}{Before Feedback} &
  \multicolumn{2}{c}{After Feedback} \\ \cline{2-5} 
\multirow{-2}{*}{\cellcolor[HTML]{FFFFFF}{\color[HTML]{333333} Student ID}} &
  CA (\%)&
  CAD (\%)&
  CA (\%)&
  CAD (\%)\\ \hline
\rowcolor[HTML]{FFF2CC} 
\cellcolor[HTML]{FFFFFF}{\color[HTML]{333333} stu7170} &
  38.62 &
  51.33 &
  52.15 &
  68.79 \\
\cellcolor[HTML]{FFFFFF}{\color[HTML]{333333} stu7123} &
  \cellcolor[HTML]{D9E1F2}36.47 &
  \cellcolor[HTML]{D9E1F2}49.15 &
  \cellcolor[HTML]{FFF2CC}49.82 &
  \cellcolor[HTML]{FFF2CC}66.38 \\
\cellcolor[HTML]{FFFFFF}{\color[HTML]{333333} stu3006} &
  \cellcolor[HTML]{D9E1F2}52.89 &
  \cellcolor[HTML]{D9E1F2}66.72 &
  \cellcolor[HTML]{FFF2CC}58.45 &
  \cellcolor[HTML]{FFF2CC}73.19 \\
\cellcolor[HTML]{FFFFFF}{\color[HTML]{333333} stu4015} &
  \cellcolor[HTML]{FFF2CC}58.93 &
  \cellcolor[HTML]{FFF2CC}68.25 &
  \cellcolor[HTML]{D9E1F2}54.17 &
  \cellcolor[HTML]{D9E1F2}74.51 \\
\rowcolor[HTML]{D9E1F2} 
\cellcolor[HTML]{FFFFFF}{\color[HTML]{333333} stu7218} &
  68.35 &
  80.17 &
  71.92 &
  84.03 \\
\cellcolor[HTML]{FFFFFF}{\color[HTML]{333333} stu3006} &
  \cellcolor[HTML]{FFF2CC}71.29 &
  \cellcolor[HTML]{FFF2CC}79.41 &
  \cellcolor[HTML]{D9E1F2}67.83 &
  \cellcolor[HTML]{D9E1F2}83.66 \\ \hline
\end{tabular}%
\caption{Performance changes of two student groups before and after instructional feedback and optimization in a real class, with identical background colors indicating group membership. CA and CAD denote Core Action Proportion and Core Action Duration Proportion, respectively. The case visualization is in the Appendix.}
\label{real-case}
\end{table}

Table \ref{tab:real} compares key indicators before and after teaching optimization and shows significant improvements across all metrics. Increases of 4.40\% in core action proportion and 5.46\% in duration proportion indicate that targeted grouping reduced ineffective movements and increased focus on key technical practice. Decreases in variation coefficients (22.25\% for core actions and 16.56\% for total duration) demonstrate that homogeneous grouping narrowed participation gaps, improving skill balance and equity in class resource allocation. Overall, the experiment confirms the framework’s ability to transform raw motion data into actionable teaching insights, providing strong empirical support for its use in PE instruction.

Table~\ref{real-case} shows performance changes in two student groups before and after instructional feedback and optimization in a real class. The optimization strategy regrouped students during training based on training intensity and model-generated recommendations, with identical background colors indicating group membership. The model’s recommendation component formed groups according to homogeneity, tempo alignment, and similar skill levels, assigning students with comparable abilities together. This approach aligns with established theories of cooperative learning and homogeneous training in physical education \cite{dysonUsingCooperativeLearning2001,colakhodzicIdentifyingHomogenousGroups2012,weiTestingEffectsPlayer2025} and the instructional results confirm its effectiveness in enhancing student engagement.

%% file: src/7_conclusion.tex
\section{Conclusion}

In this paper, we introduce a student behavior analysis framework tailored for physical education classes. This framework analyzes student behaviors based on motion signals collected during class and automatically generates pedagogically meaningful reports. By incorporating motion detection, activity recognition, and action quality assessment, we achieve fine-grained analysis of student actions. The integration of structured student behavior sequence data with large language models allows for the automatic generation of analysis reports at both the individual student level and the class level. Experimental results and empirical studies demonstrate that our approach offers a novel and effective method for analyzing student behaviors in physical education settings, enhancing the understanding of student engagement and instructional effectiveness, and providing valuable feedback for both teachers and students.

\clearpage
\section*{Ethical Statements}
We have obtained authorization to collect and use signals for research purposes. All identifiable information, such as names or markers, has been anonymized. The signals are strictly used for the objectives outlined in this study, with no alternative purposes. Extensive security measures have been implemented to safeguard the database from unauthorized access, ensuring responsible data use. Additionally, we support ongoing ethical reviews and community consultations to continually refine the ethical and implementation guidelines of the framework.

%% file: src/appendix.tex
\section{Appendix}

\subsection{Prompt Template Used for Report Generation}

\begin{figure*}[!t]
    \centering
    \includegraphics[width=0.85\linewidth]{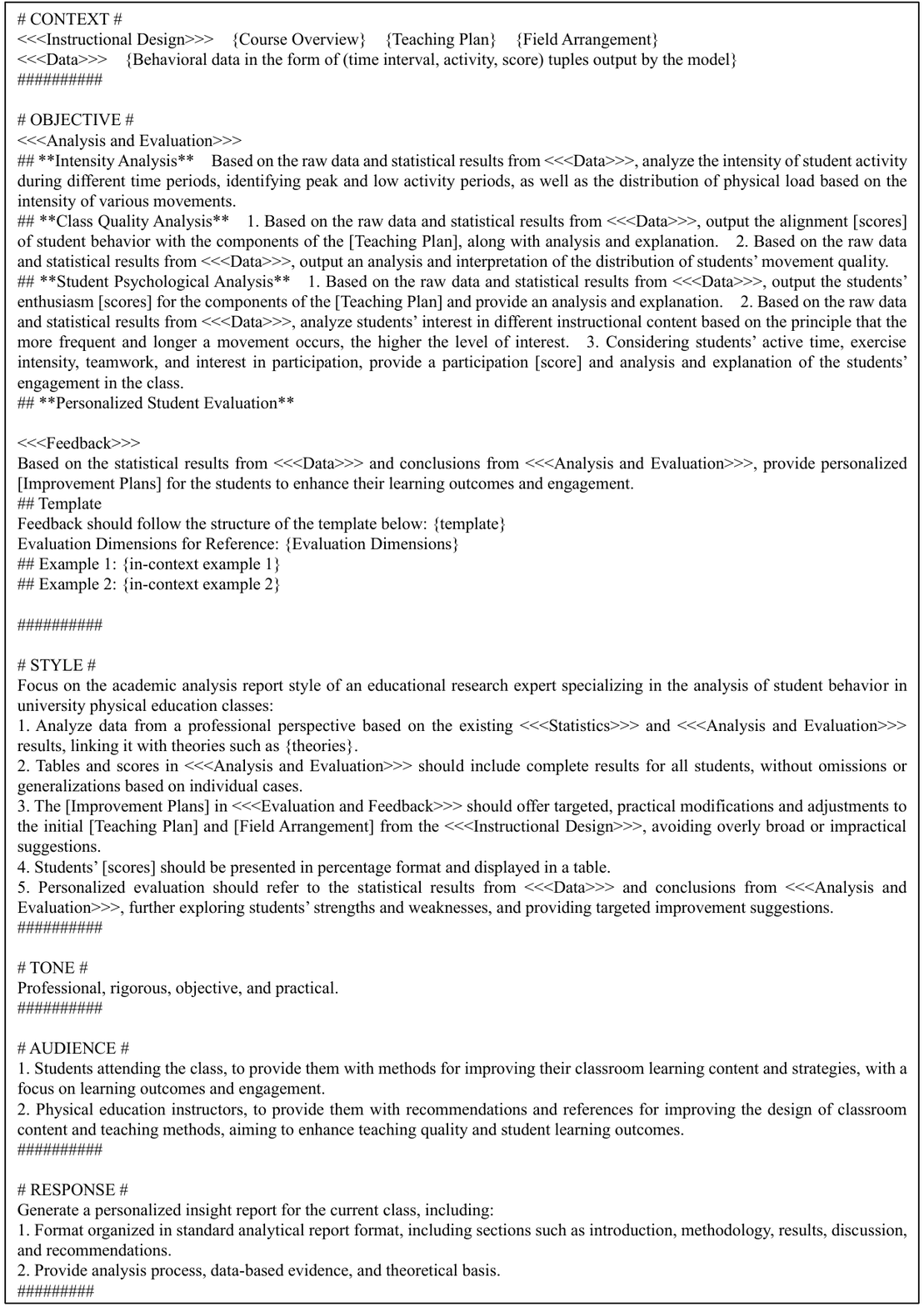}
    \caption{The prompt template we used for report generation.}
    \label{fig:template}
\end{figure*}

Figure \ref{fig:template} illustrates the prompt template we designed and refined through iterative processes. The prompt provided to the large model includes the following elements:
\begin{itemize}
    \item \textbf{Context}: Information supplied to the model, which encompasses the teacher’s instructional design, course details, and the structured triplet data output from the MD-AR-AQA model.
    \item \textbf{Objective}: The task definition, outlines the model's goals, including analyzing and evaluating student behavior based on the given data, as well as providing feedback in the format of the contextual examples.
    \item \textbf{Style} and \textbf{Tone}: A detailed description of the required format and language style for the model's output.
    \item \textbf{Audience}: The target audience for the report, with the model generating a report tailored to the specified group.
    \item \textbf{Response}: The content that the model is expected to generate.
    
\end{itemize}

The model is tasked with using the input data to conduct analysis and generate reports following the requirements and examples outlined in the prompt.

\subsection{Additional Details about Case Study}

\begin{figure}[!t]
    \centering
    \includegraphics[width=\linewidth]{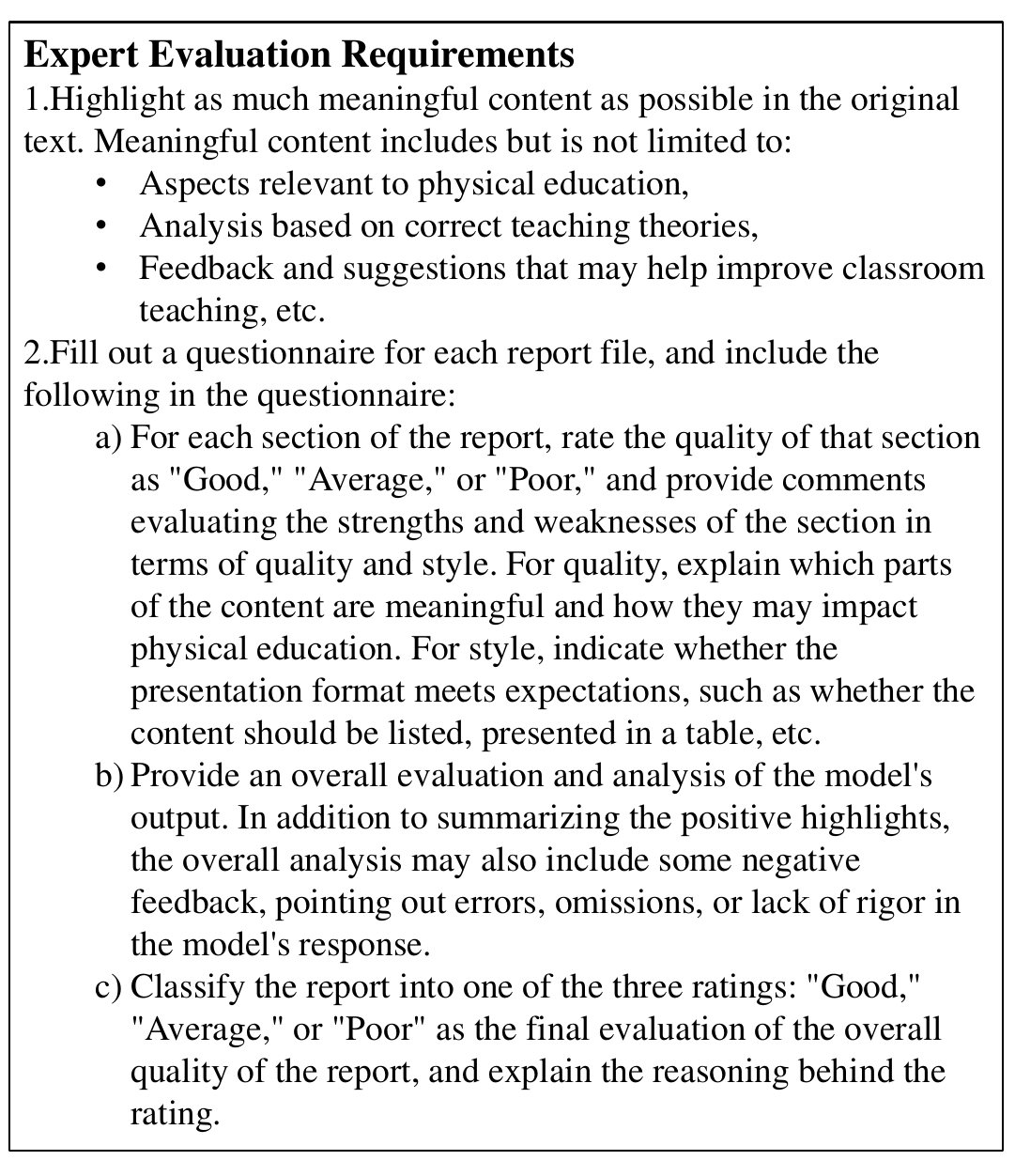}
    \caption{The instructions we used to perform the expert evaluation of the generated reports.}
    \label{fig:instuction}
\end{figure}

Figure \ref{fig:instuction}  presents the instructions provided to human experts for conducting manual evaluations. The experts were instructed to highlight meaningful content within the original report and complete a questionnaire assessing each section.

Since not all evaluators have taught the same basketball class, and the teaching content, process, and training intensity may vary between different classes, we did not have all evaluators assess all reports. Instead, we allocated the reports in the following way to ensure that each evaluator fully understood the course content, teaching process, and training intensity involved in the report they were evaluating: We assigned evaluators to the reports, ensuring that each report was evaluated by at least two evaluators (as described in Section 6 of our paper). These evaluators were all present for the specific class that the report pertained to, allowing them to gain a comprehensive understanding of the course content, teaching process, and training intensity. The evaluation of the reports was based on the consensus of the evaluators. In cases where there were disagreements among evaluators, we invited an additional evaluator—who either participated in or closely watched the video of the class—to make the final judgment.

\begin{figure}[!t]
    \centering
    \includegraphics[width=\linewidth]{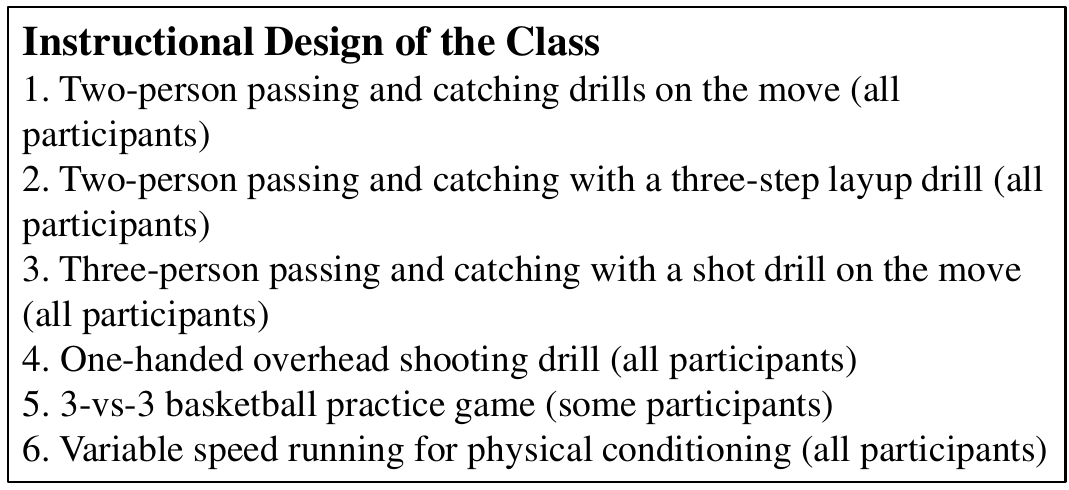}
    \caption{The instructional plan utilized in the classes.}
    \label{fig:class-design}
\end{figure}

Figure \ref{fig:class-design} illustrates the instructional plan utilized in the class during the empirical study. Both individual and class reports effectively demonstrate the relevance of the analytical process to the instructional plan, indicating that the motion detection, activity recognition, and action quality analysis models employed before report generation are able to accurately identify actions within the signal, which is the basis for generating high-quality reports.

\begin{figure*}[!t]
    \centering
    \includegraphics[width=\linewidth]{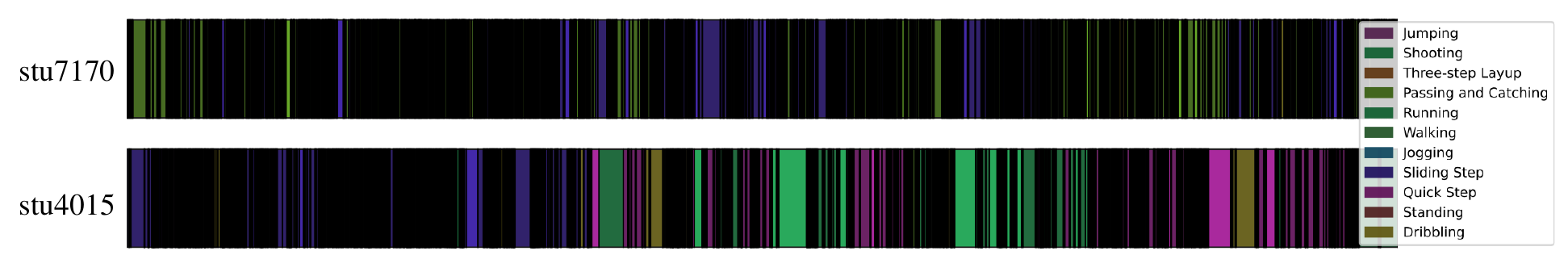}
    \caption{The visualization of student behaviors in the real-world physical education experiment before optimization.}
    \label{fig:real-before}
\end{figure*}

\begin{figure*}[!t]
    \centering
    \includegraphics[width=\linewidth]{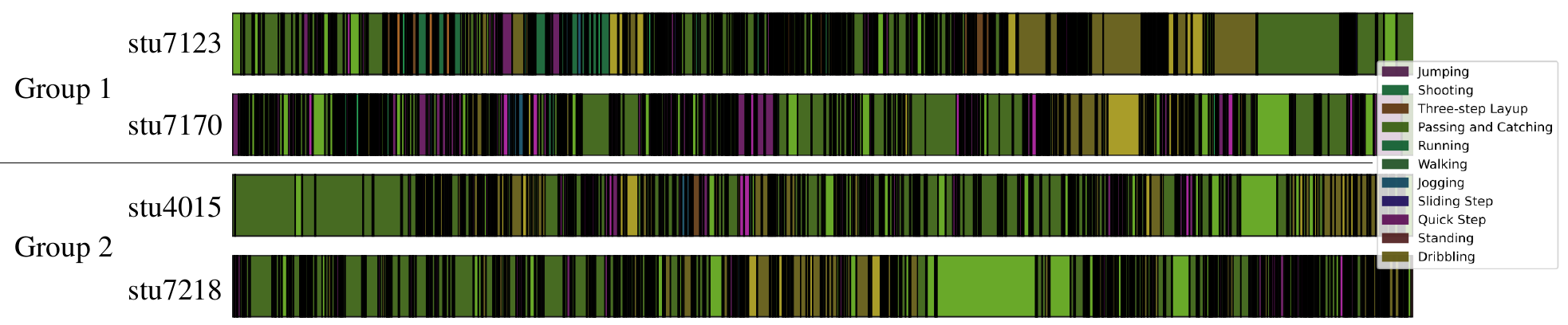}
    \caption{The visualization of student behaviors in the real-world physical education experiment after optimization.}
    \label{fig:real-after}
\end{figure*}

\subsection{Visualization of the Real Class Experiment}

Figures~\ref{fig:real-before} and~\ref{fig:real-after} respectively illustrate the visualization of student behaviors in the real-world physical education experiment before and after classroom optimization. Prior to optimization, stu7170 and stu4015 were assigned to the same practice group. However, substantial disparities in skill level and engagement resulted in poor movement consistency and limited coordination during group exercises. Following instructional feedback and regrouping, each student was matched with peers of comparable ability, leading to a marked improvement in both participation levels and action consistency. These results indicate that the proposed grouping strategy can effectively enhance students’ engagement, learning interest, and overall instructional outcomes.

\subsection{More Examples of Reports and Expert Evaluations}

\begin{figure*}[!t]
    \centering
    \includegraphics[width=0.9\linewidth]{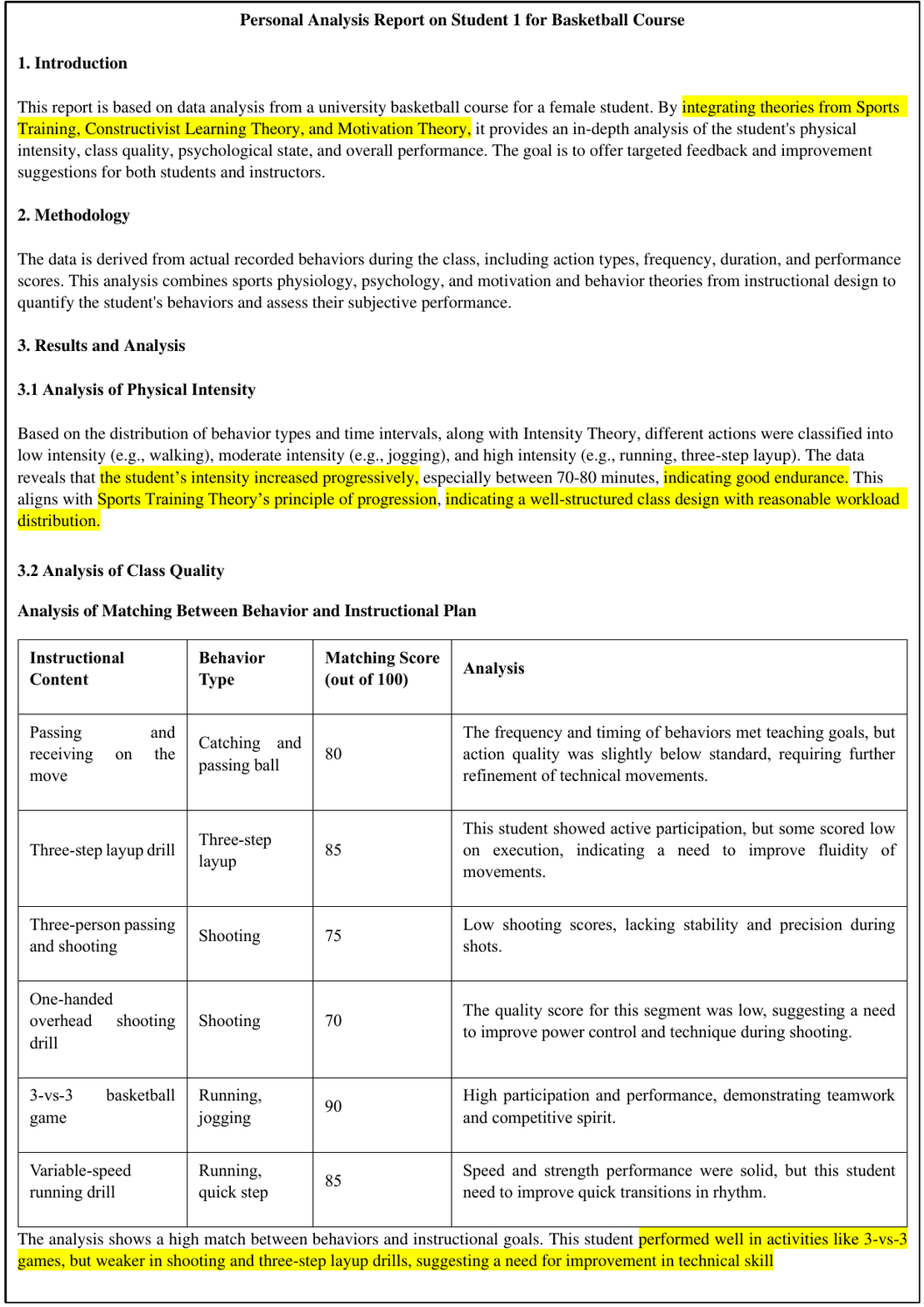}
    \caption{The complete personal report, part 1 out of 2.}
    \label{fig:individual-report-pg1}
\end{figure*}

\begin{figure*}[!t]
    \centering
    \includegraphics[width=0.9\linewidth]{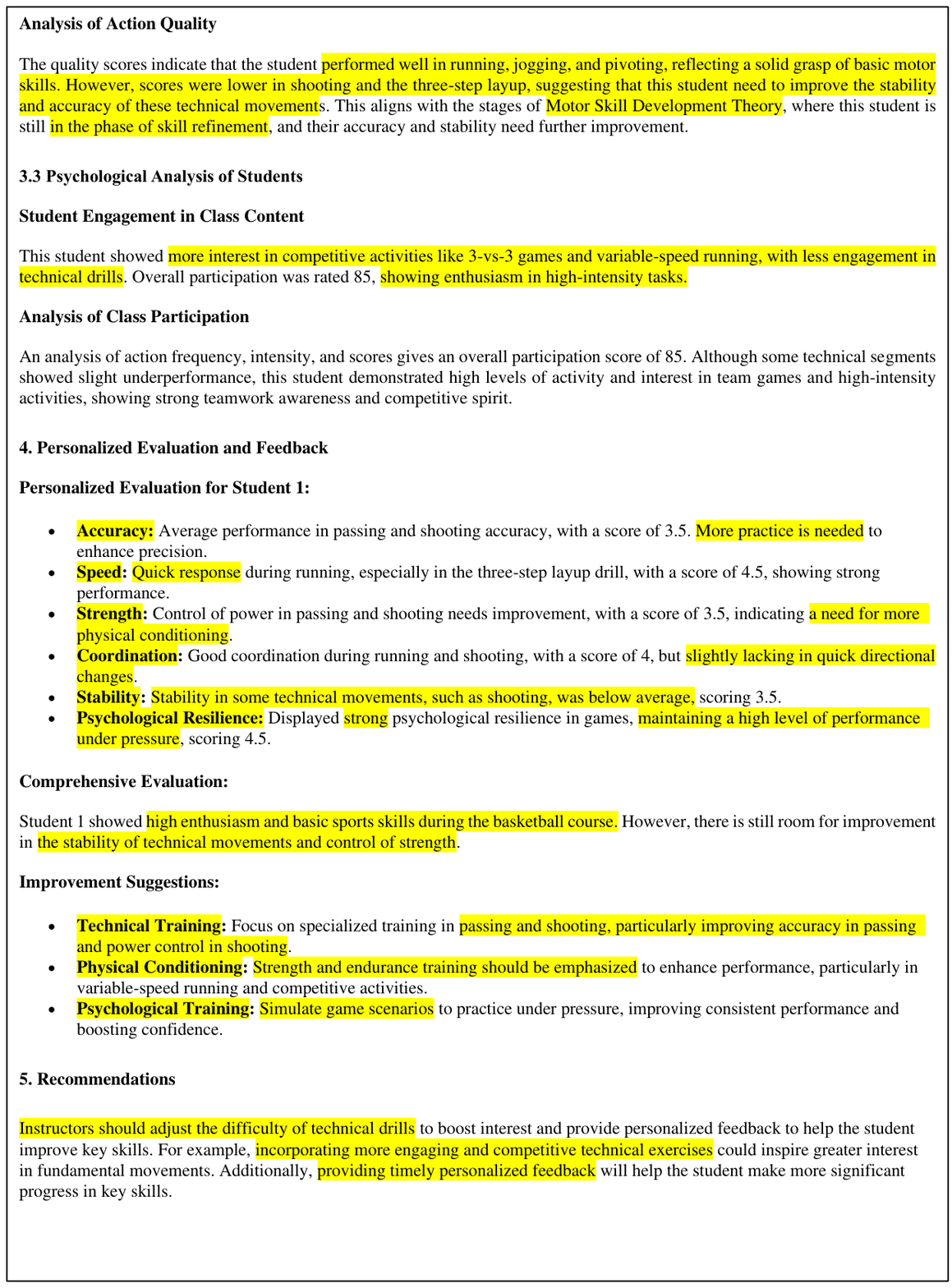}
    \caption{The complete personal report, part 2 out of 2.}
    \label{fig:individual-report-pg2}
\end{figure*}

\begin{figure*}[!t]
    \centering
    \includegraphics[width=0.85\linewidth]{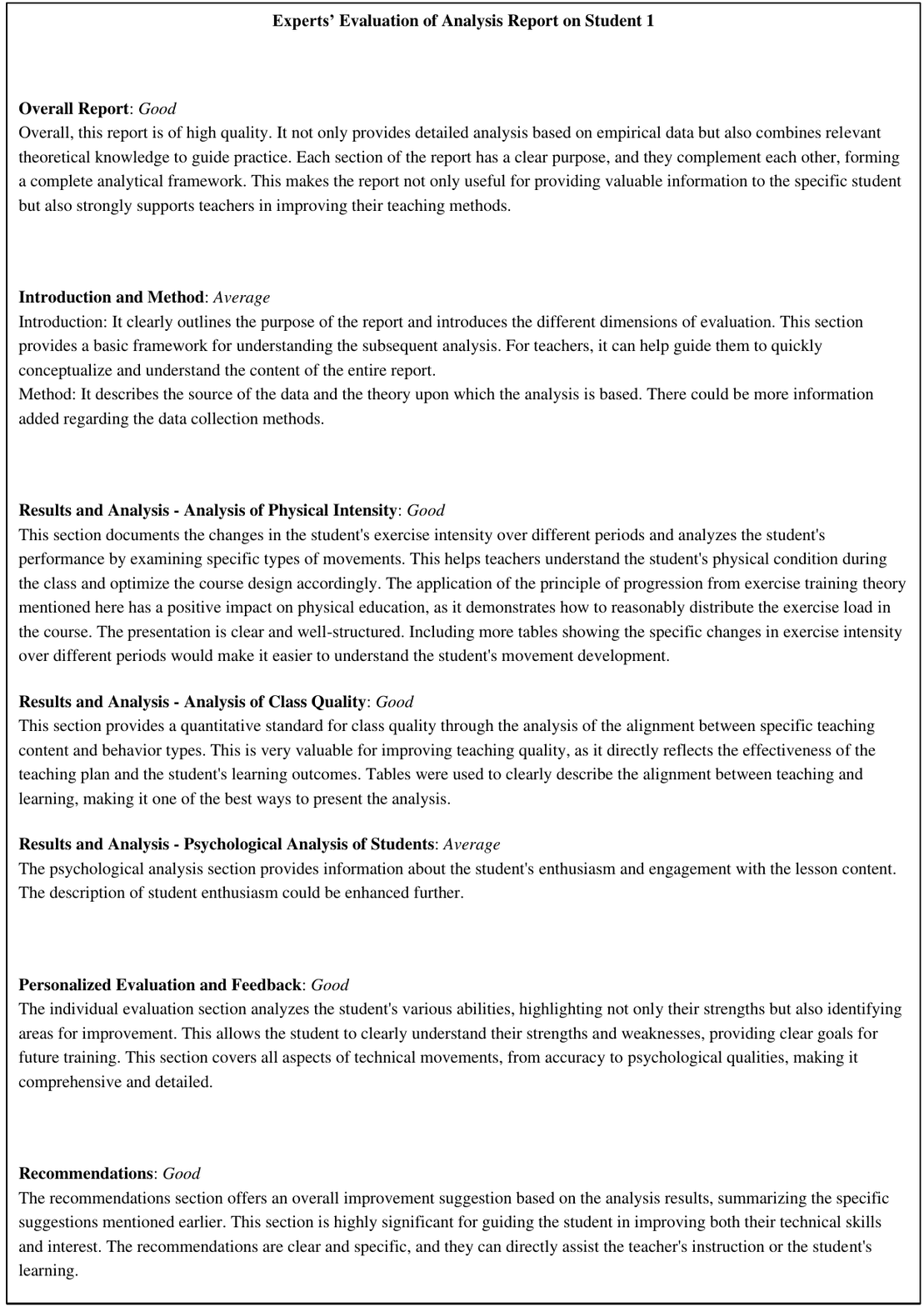}
    \caption{Expert evaluation for the personal report.}
    \label{fig:student-report-expert-score}
\end{figure*}

Figure \ref{fig:individual-report-pg1} and \ref{fig:individual-report-pg2} display a comprehensive analysis report for an individual student, while Figure \ref{fig:student-report-expert-score} shows the expert's evaluation of the report. The expert deemed the overall quality of the report to be high, as it not only provides a detailed analysis based on empirical data but also integrates relevant theoretical knowledge to inform practice. The report offers valuable insights for the specific student and provides strong support for teachers to improve their instructional methods.

\begin{figure*}[!t]
    \centering
    \includegraphics[width=0.9\linewidth]{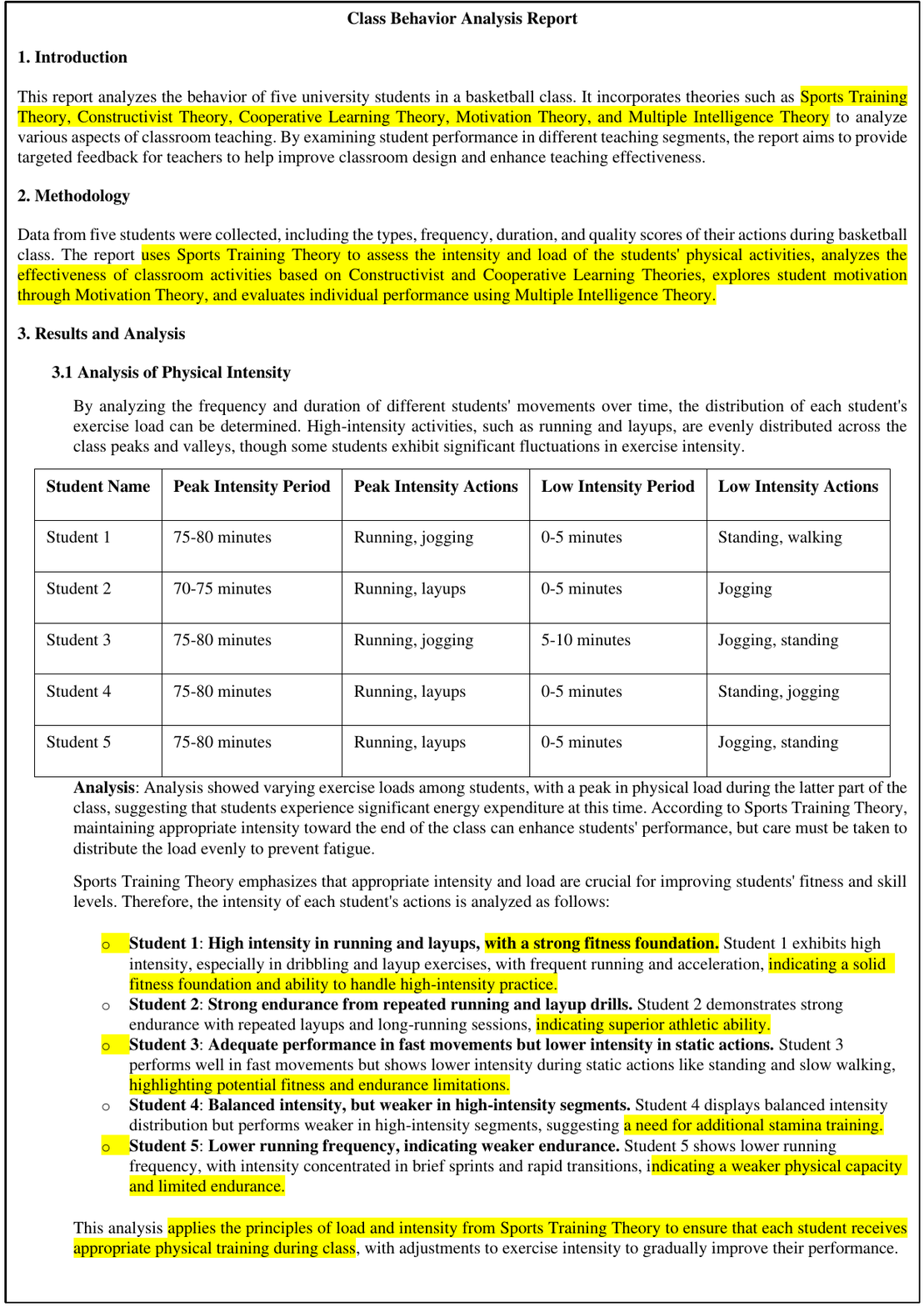}
    \caption{The complete class report, part 1 out of 4.}
    \label{fig:class-report-1-5-pg1}
\end{figure*}

\begin{figure*}[!t]
    \centering
    \includegraphics[width=0.9\linewidth]{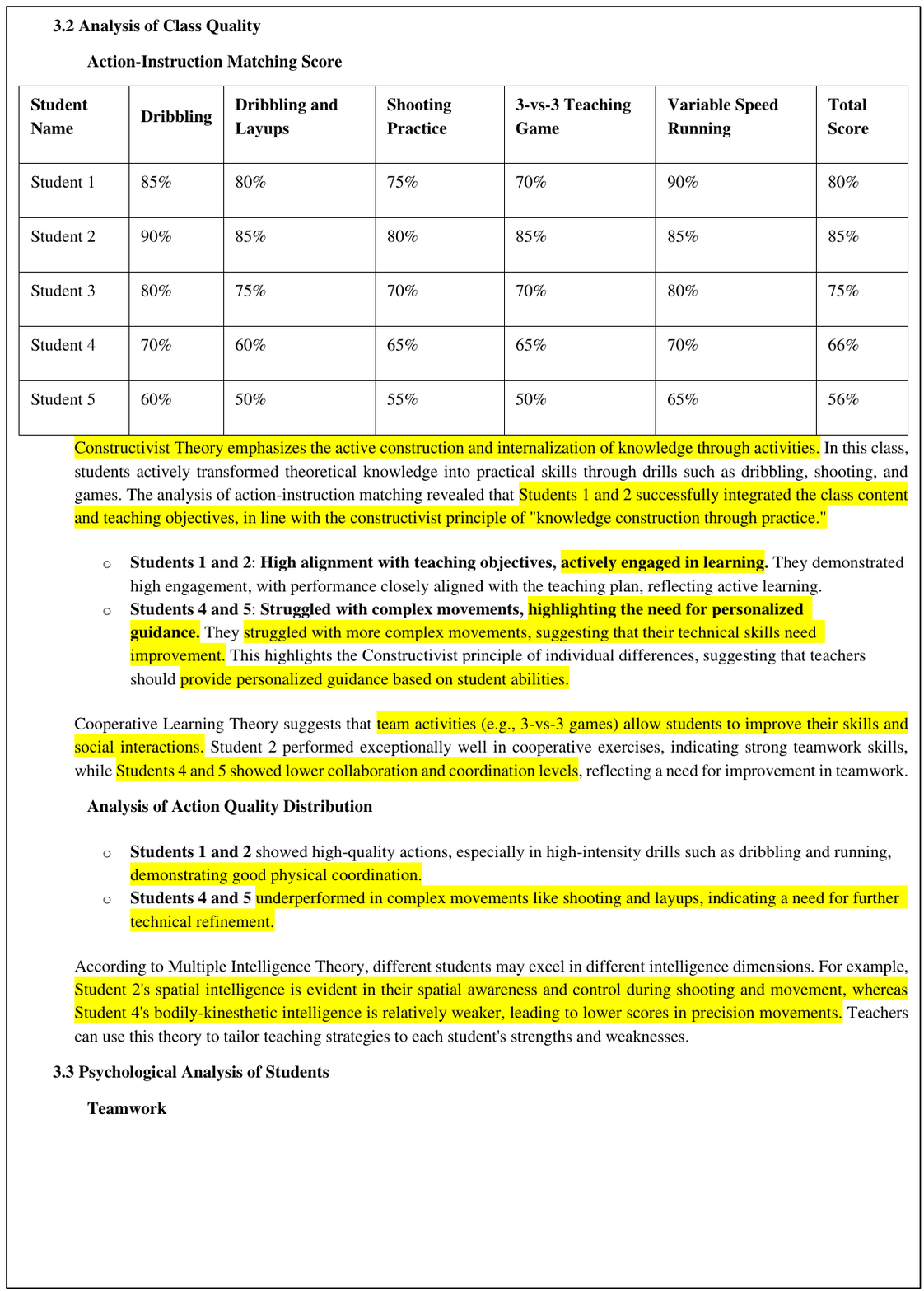}
    \caption{The complete class report, part 2 out of 4.}
    \label{fig:class-report-1-5-pg2}
\end{figure*}

\begin{figure*}[!t]
    \centering
    \includegraphics[width=0.9\linewidth]{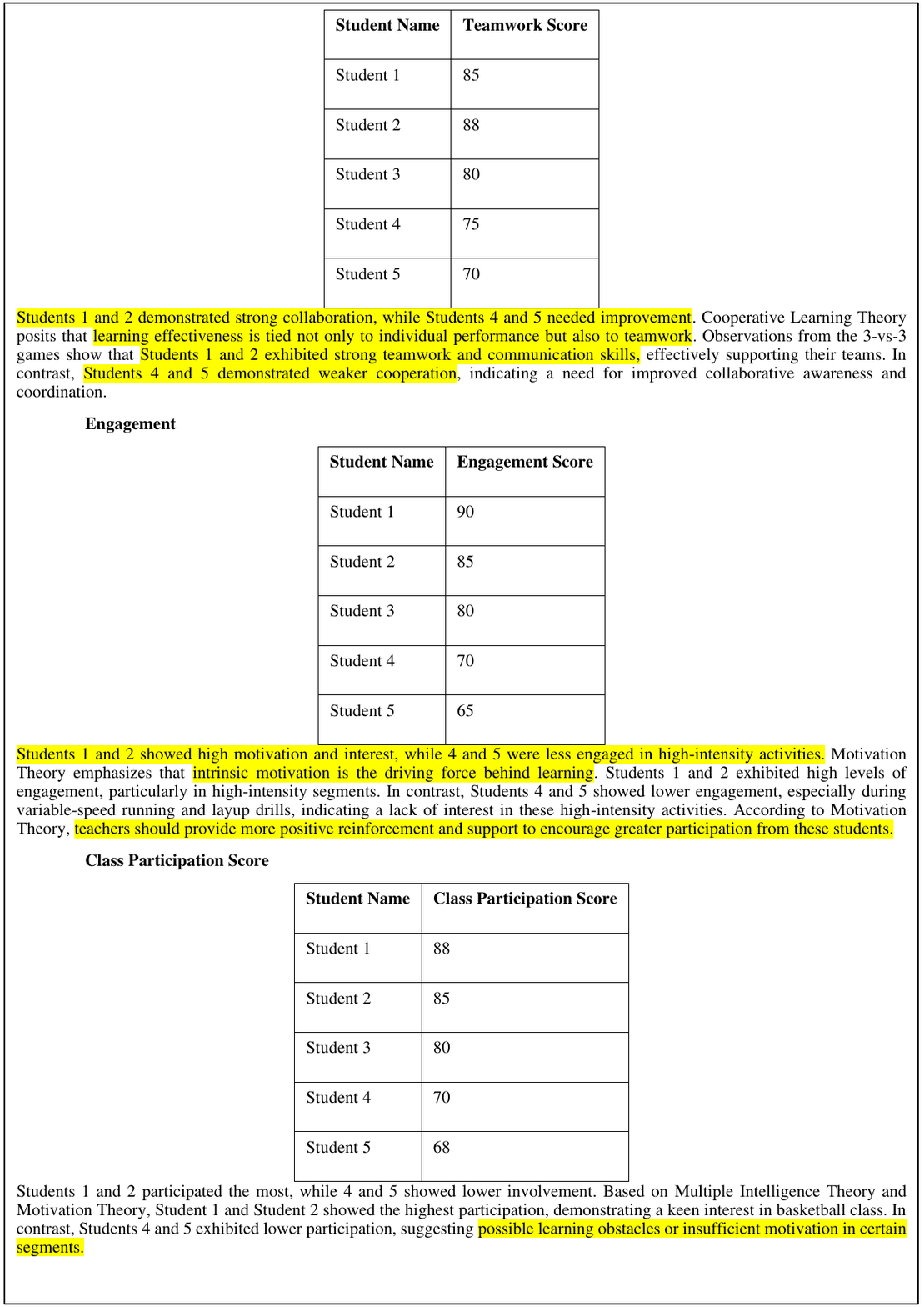}
    \caption{The complete class report, part 3 out of 4.}
    \label{fig:class-report-1-5-pg3}
\end{figure*}

\begin{figure*}[!t]
    \centering
    \includegraphics[width=0.9\linewidth]{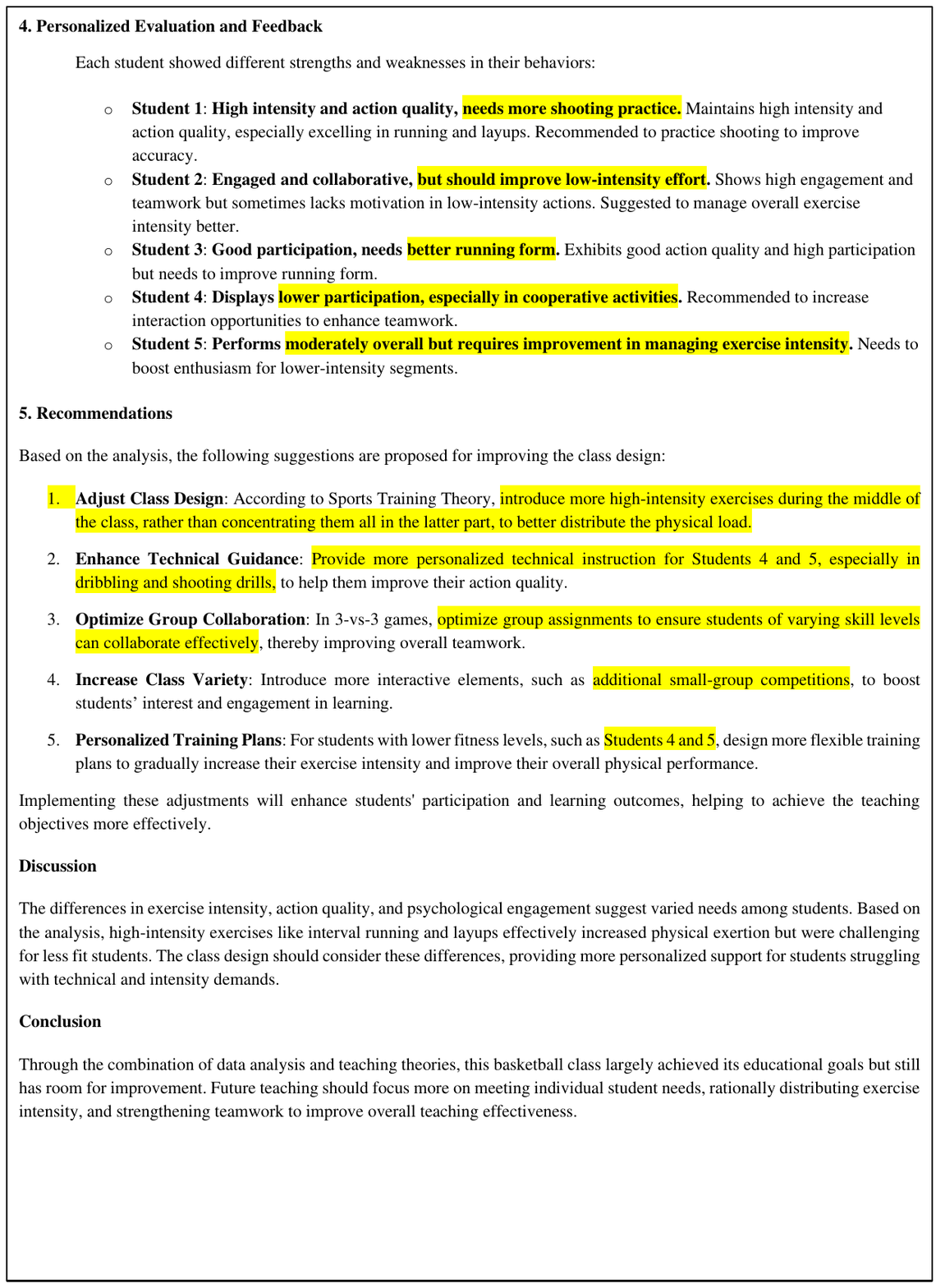}
    \caption{The complete class report, part 4 out of 4.}
    \label{fig:class-report-1-5-pg4}
\end{figure*}

\begin{figure*}[!t]
    \centering
    \includegraphics[width=0.85\linewidth]{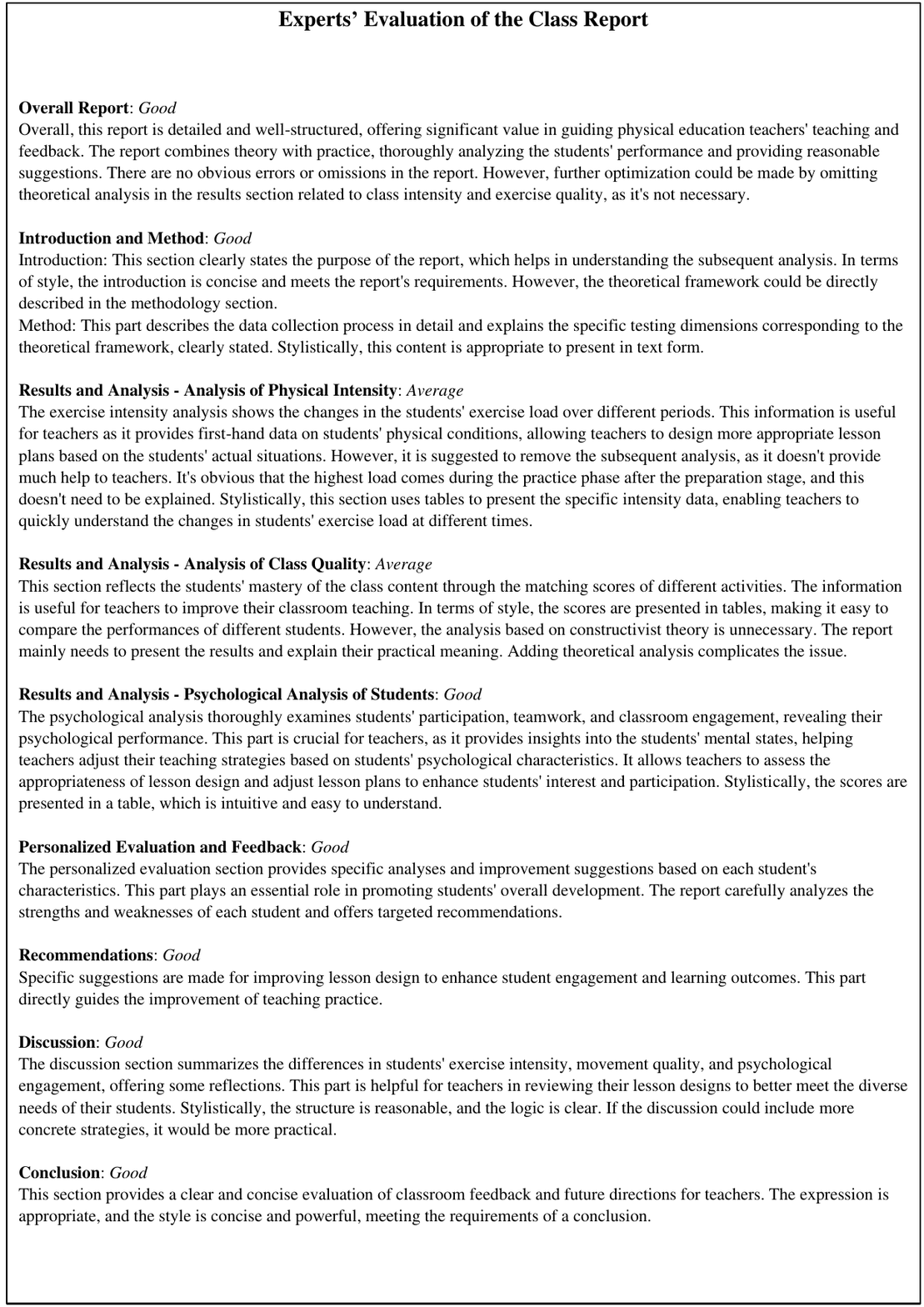}
    \caption{Expert evaluation for the class report.}
    \label{fig:class-report-expert-score}
\end{figure*}

Figures \ref{fig:class-report-1-5-pg1}, \ref{fig:class-report-1-5-pg2}, \ref{fig:class-report-1-5-pg3}, and \ref{fig:class-report-1-5-pg4} represent a complete analysis report for an entire class, with Figure \ref{fig:class-report-expert-score} reflecting the expert's evaluation. The expert found this report to be highly valuable for guiding physical education teachers in their teaching and feedback processes. Additionally, the expert offered suggestions for improvement, focusing on areas of greater concern to teachers, such as placing less emphasis on theoretical content and more on feasible, practical solutions. The report could appropriately reduce the proportion of theoretical content.

\begin{figure*}[!t]
    \centering
    \includegraphics[width=0.9\linewidth]{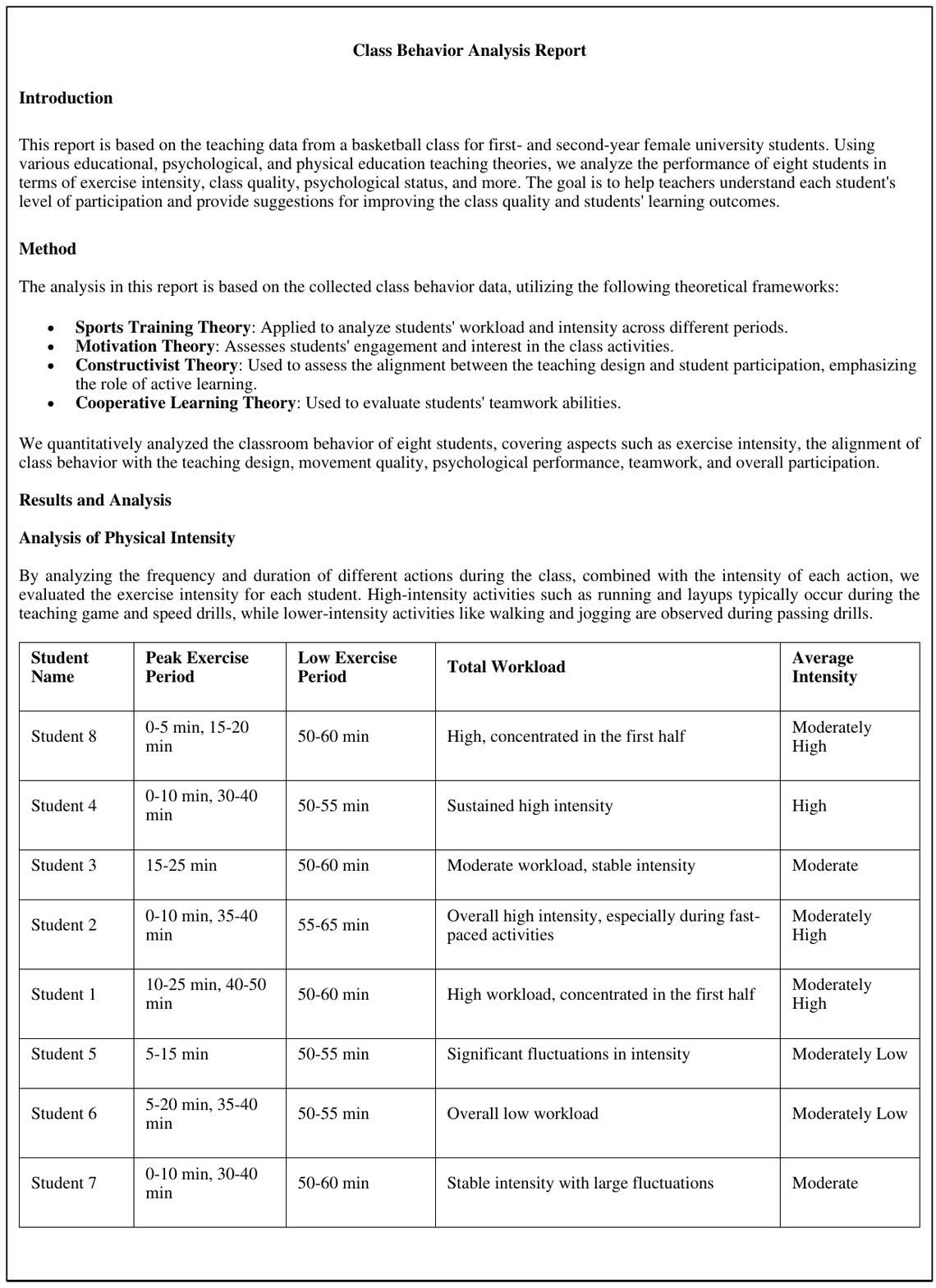}
    \caption{The complete report for another class, part 1 out of 4.}
    \label{fig:class-report-pg1}
\end{figure*}

\begin{figure*}[!t]
    \centering
    \includegraphics[width=0.9\linewidth]{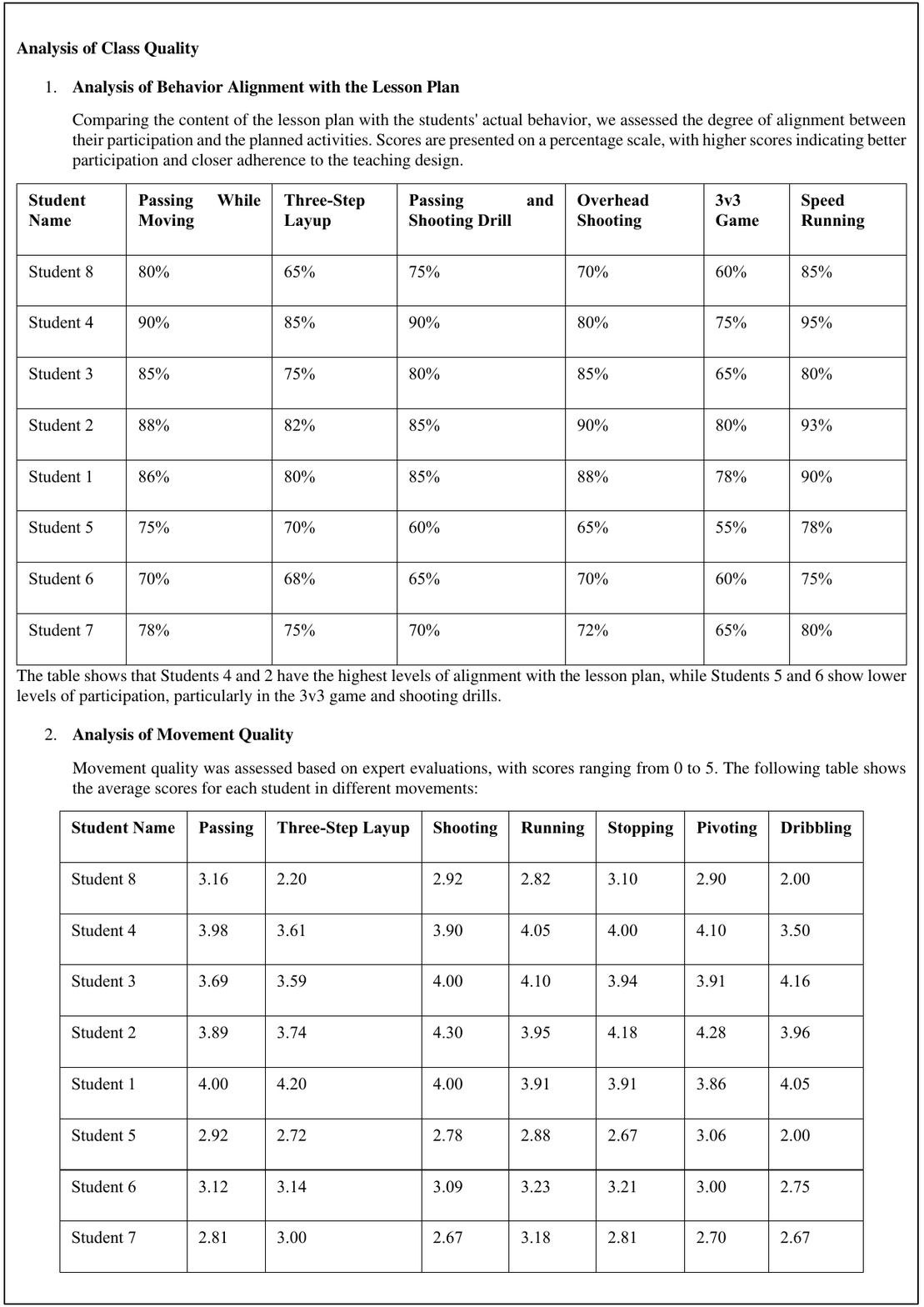}
    \caption{The complete report for another class, part 2 out of 4.}
    \label{fig:class-report-pg2}
\end{figure*}

\begin{figure*}[!t]
    \centering
    \includegraphics[width=0.9\linewidth]{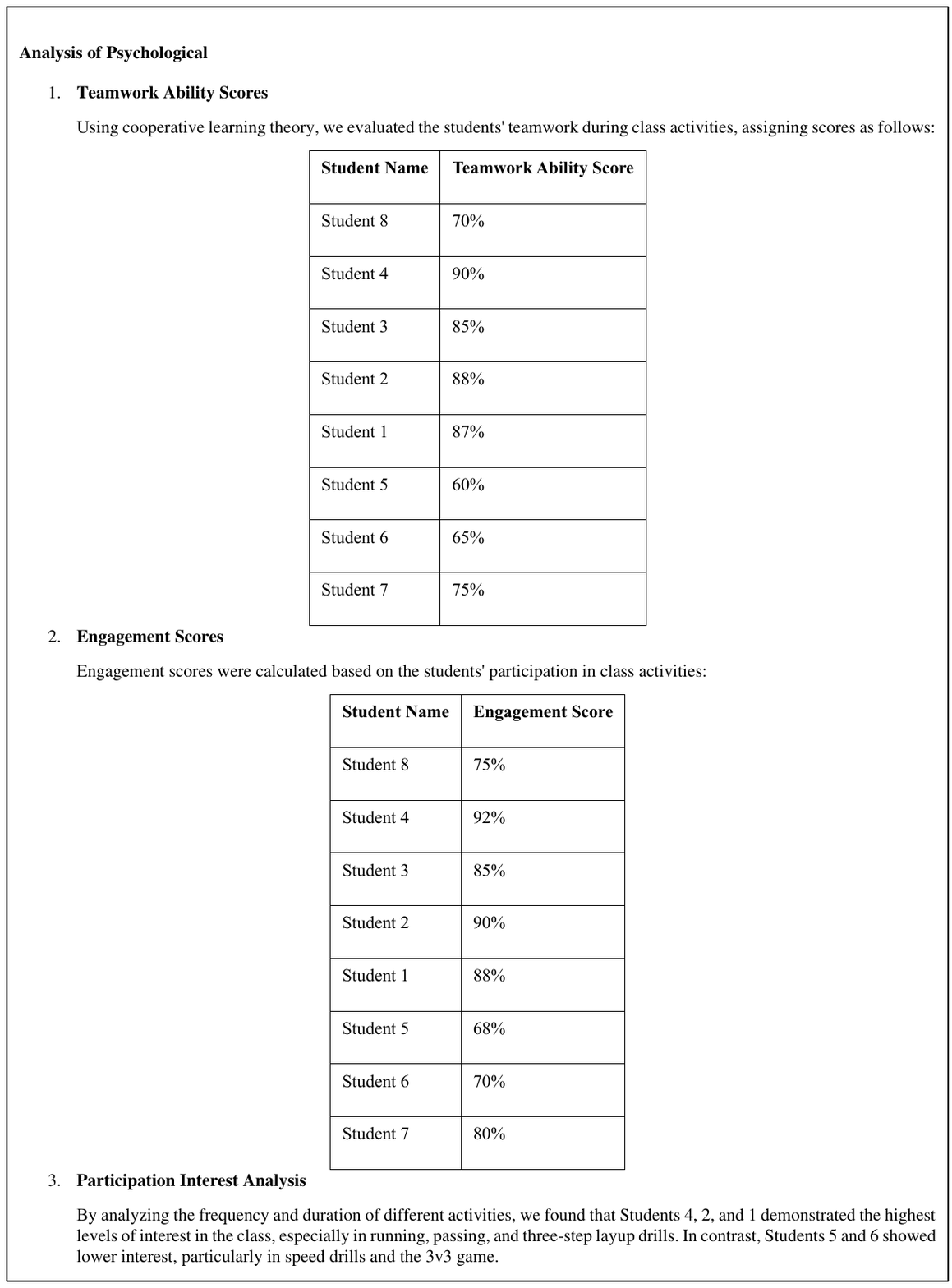}
    \caption{The complete report for another class, part 3 out of 4.}
    \label{fig:class-report-pg3}
\end{figure*}

\begin{figure*}[!t]
    \centering
    \includegraphics[width=0.9\linewidth]{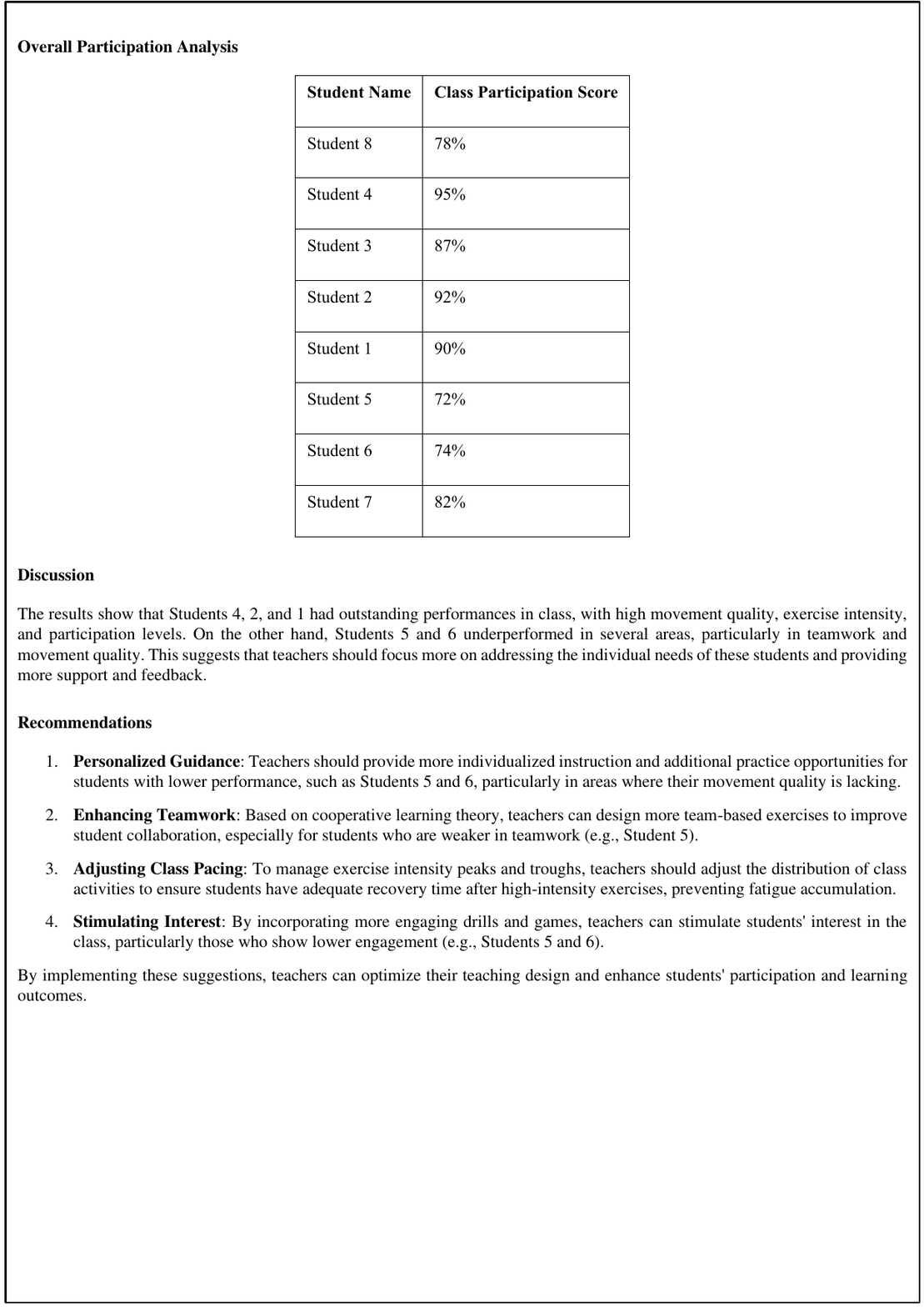}
    \caption{The complete report for another class, part 4 out of 4.}
    \label{fig:class-report-pg4}
\end{figure*}

\begin{figure*}[!t]
    \centering
    \includegraphics[width=0.9\linewidth]{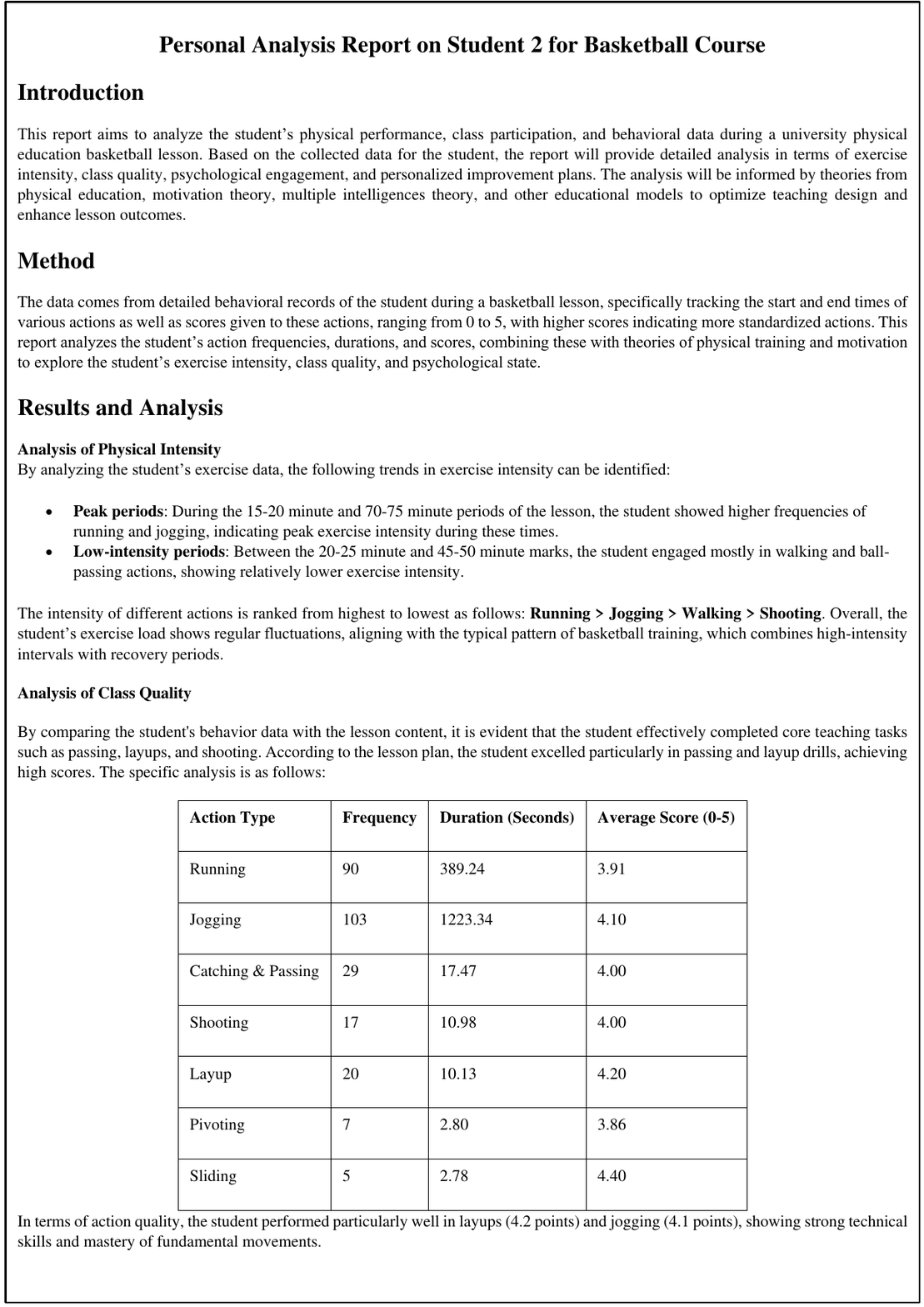}
    \caption{The complete report for a student in another class, part 1 out of 2.}
    \label{fig:individual-report-1-pg1}
\end{figure*}

\begin{figure*}[!t]
    \centering
    \includegraphics[width=0.9\linewidth]{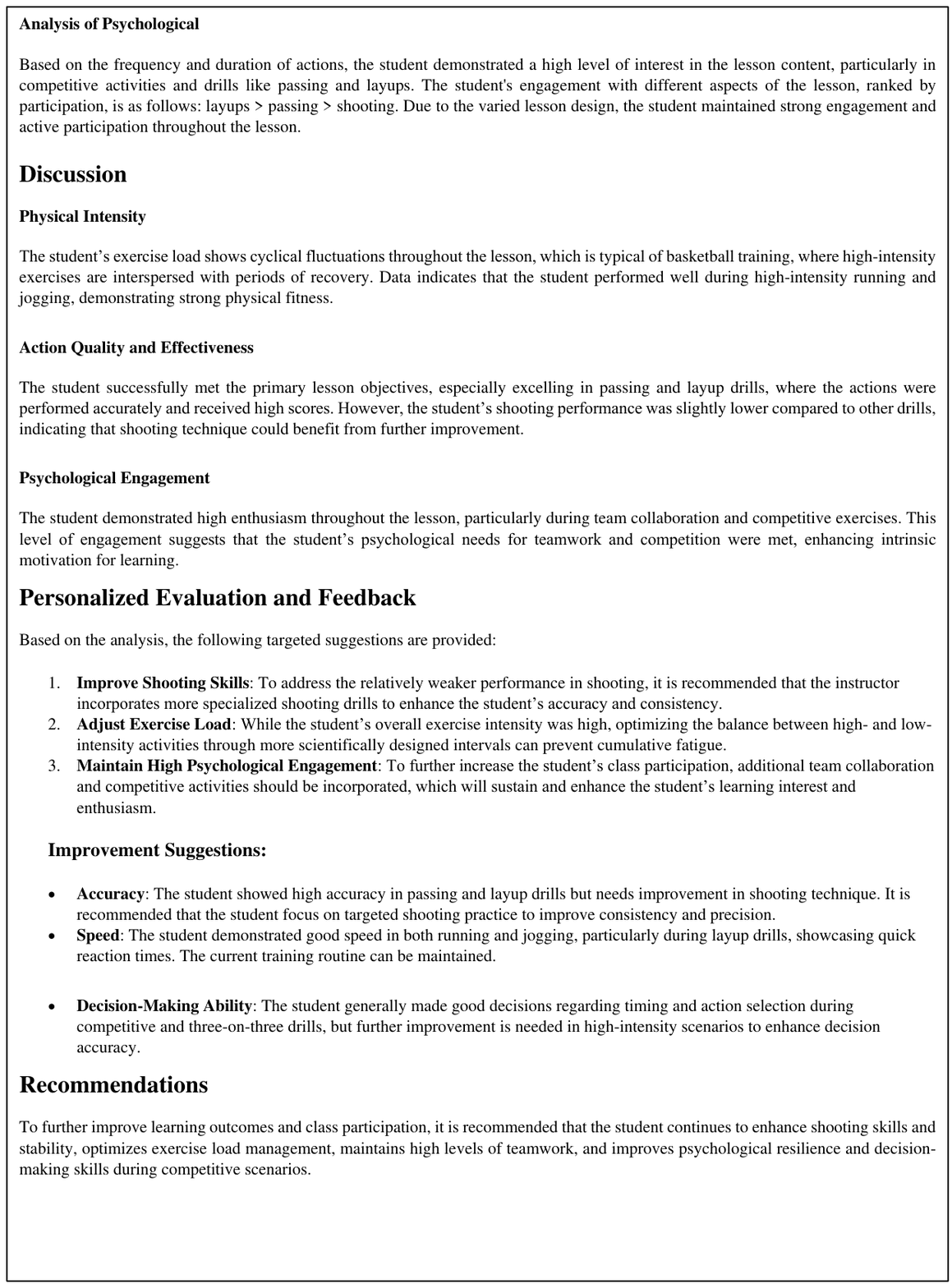}
    \caption{The complete report for a student in another class, part 2 out of 2.}
    \label{fig:individual-report-1-pg2}
\end{figure*}

To further validate the effectiveness of our approach, we conduct analysis on another class and present the class report in Figures \ref{fig:class-report-pg1}, \ref{fig:class-report-pg2}, \ref{fig:class-report-pg3}, and \ref{fig:class-report-pg4} and a personal report in Figures \ref{fig:individual-report-1-pg1} and \ref{fig:individual-report-1-pg2}. A comparison of reports from different classes reveals that our framework maintains consistency in both report format and analytical approach. Despite the larger student cohort in this class, the model was able to accurately capture the relationships and differences among a greater number of students, thus demonstrating the robustness and applicability of our framework across diverse scenarios.

\subsection{Limitations and Future Works}

The action recognition and quality assessment models discussed in this paper require a small number of sample actions for fine-tuning. While sufficient samples can be collected in practice, the diversity of physical education scenarios and the lack of context-specific data limit full generalization without fine-tuning. To address this, future work will explore leveraging the generative capabilities of large language models through multi-agent interaction to generate motion signals in physical education contexts. By simulating and generating a large volume of reliable motion signals, we aim to overcome data scarcity and improve model generalization.

%% file: ijcai26.bib
@article{anguitaPublicDomainDataset2013,
  title = {A {{Public Domain Dataset}} for {{Human Activity Recognition Using Smartphones}}},
  author = {Anguita, Davide and Ghio, Alessandro and Oneto, Luca and Parra, Xavier and {Reyes-Ortiz}, Jorge L},
  year = {2013},
  journal = {Computational Intelligence}
}

@article{ashryCHARMDeepContinuousHuman2020,
  title = {{{CHARM-Deep}}: {{Continuous Human Activity Recognition Model Based}} on {{Deep Neural Network Using IMU Sensors}} of {{Smartwatch}}},
  shorttitle = {{{CHARM-Deep}}},
  author = {Ashry, Sara and Ogawa, Tetsuji and Gomaa, Walid},
  year = {2020},
  month = aug,
  journal = {IEEE Sensors Journal},
  volume = {20},
  number = {15},
  pages = {8757--8770},
  doi = {10.1109/JSEN.2020.2985374},
  urldate = {2024-09-03}
}

@article{baileyPhysicalEducationSport2006,
  title = {Physical {{Education}} and {{Sport}} in {{Schools}}: {{A Review}} of {{Benefits}} and {{Outcomes}}},
  shorttitle = {Physical {{Education}} and {{Sport}} in {{Schools}}},
  author = {Bailey, Richard},
  year = {2006},
  month = oct,
  journal = {Journal of School Health},
  volume = {76},
  number = {8},
  pages = {397--401},
  doi = {10.1111/j.1746-1561.2006.00132.x},
  urldate = {2024-09-02}
}

@misc{chenMotionLLMUnderstandingHuman2024,
  title = {{{MotionLLM}}: {{Understanding Human Behaviors}} from {{Human Motions}} and {{Videos}}},
  shorttitle = {{{MotionLLM}}},
  author = {Chen, Ling-Hao and Lu, Shunlin and Zeng, Ailing and Zhang, Hao and Wang, Benyou and Zhang, Ruimao and Zhang, Lei},
  year = {2024},
  month = may,
  number = {arXiv:2405.20340},
  eprint = {2405.20340},
  primaryclass = {cs},
  publisher = {arXiv},
  doi = {10.48550/arXiv.2405.20340},
  urldate = {2024-09-08},
  archiveprefix = {arXiv}
}

@inproceedings{chuangDebiasedContrastiveLearning2020,
  title = {Debiased {{Contrastive Learning}}},
  booktitle = {Advances in {{Neural Information Processing Systems}}},
  author = {Chuang, Ching-Yao and Robinson, Joshua and Lin, Yen-Chen and Torralba, Antonio and Jegelka, Stefanie},
  year = {2020},
  volume = {33},
  pages = {8765--8775},
  publisher = {Curran Associates, Inc.},
  urldate = {2024-09-14}
}

@article{deldariCOCOACrossModality2022,
  title = {{{COCOA}}: {{Cross Modality Contrastive Learning}} for {{Sensor Data}}},
  shorttitle = {{{COCOA}}},
  author = {Deldari, Shohreh and Xue, Hao and Saeed, Aaqib and Smith, Daniel V. and Salim, Flora D.},
  year = {2022},
  month = sep,
  journal = {Proceedings of the ACM on Interactive, Mobile, Wearable and Ubiquitous Technologies},
  volume = {6},
  number = {3},
  eprint = {2208.00467},
  primaryclass = {cs},
  pages = {1--28},
  doi = {10.1145/3550316},
  urldate = {2024-09-14},
  archiveprefix = {arXiv}
}

@article{fredricksSchoolEngagementPotential2004,
  title = {School {{Engagement}}: {{Potential}} of the {{Concept}}, {{State}} of the {{Evidence}}},
  shorttitle = {School {{Engagement}}},
  author = {Fredricks, Jennifer A. and Blumenfeld, Phyllis C. and Paris, Alison H.},
  year = 2004,
  journal = {Review of Educational Research},
  volume = {74},
  number = {1},
  pages = {59--109},
  publisher = {American Educational Research Association},
  urldate = {2024-09-02}
}

@inproceedings{girdharImageBindOneEmbedding2023a,
  title = {{{ImageBind One Embedding Space}} to {{Bind Them All}}},
  booktitle = {2023 {{IEEE}}/{{CVF Conference}} on {{Computer Vision}} and {{Pattern Recognition}} ({{CVPR}})},
  author = {Girdhar, Rohit and {El-Nouby}, Alaaeldin and Liu, Zhuang and Singh, Mannat and Alwala, Kalyan Vasudev and Joulin, Armand and Misra, Ishan},
  year = {2023},
  month = jun,
  pages = {15180--15190},
  publisher = {IEEE},
  address = {Vancouver, BC, Canada},
  doi = {10.1109/CVPR52729.2023.01457},
  urldate = {2024-08-02}
}

@article{guanEnsemblesDeepLSTM2017,
  title = {Ensembles of {{Deep LSTM Learners}} for {{Activity Recognition}} Using {{Wearables}}},
  author = {Guan, Yu and Pl{\"o}tz, Thomas},
  year = {2017},
  month = jun,
  journal = {Proceedings of the ACM on Interactive, Mobile, Wearable and Ubiquitous Technologies},
  volume = {1},
  number = {2},
  pages = {1--28},
  doi = {10.1145/3090076},
  urldate = {2024-09-03}
}

@article{huangShallowConvolutionalNeural2021,
  title = {Shallow {{Convolutional Neural Networks}} for {{Human Activity Recognition Using Wearable Sensors}}},
  author = {Huang, Wenbo and Zhang, Lei and Gao, Wenbin and Min, Fuhong and He, Jun},
  year = {2021},
  journal = {IEEE Transactions on Instrumentation and Measurement},
  volume = {70},
  pages = {1--11},
  doi = {10.1109/TIM.2021.3091990},
  urldate = {2024-09-03}
}

@inproceedings{huLoRALowRankAdaptation2021,
  title = {{{LoRA}}: {{Low-Rank Adaptation}} of {{Large Language Models}}},
  shorttitle = {{{LoRA}}},
  booktitle = {International {{Conference}} on {{Learning Representations}}},
  author = {Hu, Edward J. and Shen, Yelong and Wallis, Phillip and {Allen-Zhu}, Zeyuan and Li, Yuanzhi and Wang, Shean and Wang, Lu and Chen, Weizhu},
  year = {2021},
  month = oct,
  urldate = {2024-09-04}
}

@article{huRealTimeClassroomBehavior2024,
  title = {Real-{{Time Classroom Behavior Analysis}} for {{Enhanced Engineering Education}}: {{An AI-Assisted Approach}}},
  shorttitle = {Real-{{Time Classroom Behavior Analysis}} for {{Enhanced Engineering Education}}},
  author = {Hu, Jia and Huang, Zhenxi and Li, Jing and Xu, Lingfeng and Zou, Yuntao},
  year = {2024},
  month = jun,
  journal = {International Journal of Computational Intelligence Systems},
  volume = {17},
  number = {1},
  pages = {167},
  doi = {10.1007/s44196-024-00572-y},
  urldate = {2024-08-26}
}

@article{inoueDeepRecurrentNeural2018,
  title = {Deep Recurrent Neural Network for Mobile Human Activity Recognition with High Throughput},
  author = {Inoue, Masaya and Inoue, Sozo and Nishida, Takeshi},
  year = {2018},
  month = jun,
  journal = {Artificial Life and Robotics},
  volume = {23},
  number = {2},
  pages = {173--185},
  doi = {10.1007/s10015-017-0422-x},
  urldate = {2024-09-03}
}

@article{liExploringArtificialIntelligence2023,
  title = {Exploring {{Artificial Intelligence}} in {{Smart Education}}: {{Real-Time Classroom Behavior Analysis}} with {{Embedded Devices}}},
  shorttitle = {Exploring {{Artificial Intelligence}} in {{Smart Education}}},
  author = {Li, Liujun and Chen, Chao Ping and Wang, Lijun and Liang, Kai and Bao, Weiyue},
  year = {2023},
  month = jan,
  journal = {Sustainability},
  volume = {15},
  number = {10},
  pages = {7940},
  publisher = {Multidisciplinary Digital Publishing Institute},
  doi = {10.3390/su15107940},
  urldate = {2024-08-28}
}

@article{liGuidingLargeLanguage2023,
  title = {Guiding {{Large Language Models}} via {{Directional Stimulus Prompting}}},
  author = {Li, Zekun and Peng, Baolin and He, Pengcheng and Galley, Michel and Gao, Jianfeng and Yan, Xifeng},
  year = {2023},
  month = dec,
  journal = {Advances in Neural Information Processing Systems},
  volume = {36},
  pages = {62630--62656},
  urldate = {2024-09-07}
}

@article{linStudentBehaviorRecognition2021,
  title = {Student {{Behavior Recognition System}} for the {{Classroom Environment Based}} on {{Skeleton Pose Estimation}} and {{Person Detection}}},
  author = {Lin, Feng-Cheng and Ngo, Huu-Huy and Dow, Chyi-Ren and Lam, Ka-Hou and Le, Hung Linh},
  year = {2021},
  month = jan,
  journal = {Sensors},
  volume = {21},
  number = {16},
  pages = {5314},
  publisher = {Multidisciplinary Digital Publishing Institute},
  doi = {10.3390/s21165314},
  urldate = {2024-09-03}
}

@inproceedings{liSleepGestureDetection2019,
  title = {Sleep {{Gesture Detection}} in {{Classroom Monitor System}}},
  booktitle = {{{ICASSP}} 2019 - 2019 {{IEEE International Conference}} on {{Acoustics}}, {{Speech}} and {{Signal Processing}} ({{ICASSP}})},
  author = {Li, Wen and Jiang, Fei and Shen, Ruimin},
  year = {2019},
  month = may,
  pages = {7640--7644},
  doi = {10.1109/ICASSP.2019.8683116},
  urldate = {2024-09-03}
}

@incollection{liuFastAccurateHandRaising2020,
  title = {Fast and {{Accurate Hand-Raising Gesture Detection}} in {{Classroom}}},
  booktitle = {Neural {{Information Processing}}},
  author = {Liu, Tao and Jiang, Fei and Shen, Ruimin},
  editor = {Yang, Haiqin and Pasupa, Kitsuchart and Leung, Andrew Chi-Sing and Kwok, James T. and Chan, Jonathan H. and King, Irwin},
  year = {2020},
  volume = {1332},
  pages = {232--239},
  publisher = {Springer International Publishing},
  address = {Cham},
  doi = {10.1007/978-3-030-63820-7_26},
  urldate = {2024-09-03}
}

@article{liuPretrainPromptPredict2023,
  title = {Pre-Train, {{Prompt}}, and {{Predict}}: {{A Systematic Survey}} of {{Prompting Methods}} in {{Natural Language Processing}}},
  shorttitle = {Pre-Train, {{Prompt}}, and {{Predict}}},
  author = {Liu, Pengfei and Yuan, Weizhe and Fu, Jinlan and Jiang, Zhengbao and Hayashi, Hiroaki and Neubig, Graham},
  year = {2023},
  month = sep,
  journal = {ACM Computing Surveys},
  volume = {55},
  number = {9},
  pages = {1--35},
  doi = {10.1145/3560815},
  urldate = {2024-09-07}
}

@misc{malekzadehPrivacyUtilityPreserving2019,
  title = {Privacy and {{Utility Preserving Sensor-Data Transformations}}},
  author = {Malekzadeh, Mohammad and Clegg, Richard G. and Cavallaro, Andrea and Haddadi, Hamed},
  year = {2019},
  month = nov,
  number = {arXiv:1911.05996},
  eprint = {1911.05996},
  primaryclass = {cs, eess, stat},
  publisher = {arXiv},
  urldate = {2024-09-07},
  archiveprefix = {arXiv}
}

@inproceedings{malekzadehProtectingSensoryData2018,
  title = {Protecting {{Sensory Data}} against {{Sensitive Inferences}}},
  booktitle = {Proceedings of the 1st {{Workshop}} on {{Privacy}} by {{Design}} in {{Distributed Systems}}},
  author = {Malekzadeh, Mohammad and Clegg, Richard G. and Cavallaro, Andrea and Haddadi, Hamed},
  year = {2018},
  month = apr,
  pages = {1--6},
  publisher = {ACM},
  address = {Porto Portugal},
  doi = {10.1145/3195258.3195260},
  urldate = {2024-09-07}
}

@misc{moonIMU2CLIPMultimodalContrastive2022,
  title = {{{IMU2CLIP}}: {{Multimodal Contrastive Learning}} for {{IMU Motion Sensors}} from {{Egocentric Videos}} and {{Text}}},
  shorttitle = {{{IMU2CLIP}}},
  author = {Moon, Seungwhan and Madotto, Andrea and Lin, Zhaojiang and Dirafzoon, Alireza and Saraf, Aparajita and Bearman, Amy and Damavandi, Babak},
  year = {2022},
  month = oct,
  number = {arXiv:2210.14395},
  eprint = {2210.14395},
  primaryclass = {cs},
  publisher = {arXiv},
  doi = {10.48550/arXiv.2210.14395},
  urldate = {2024-09-03},
  archiveprefix = {arXiv}
}

@article{muradDeepRecurrentNeural2017,
  title = {Deep {{Recurrent Neural Networks}} for {{Human Activity Recognition}}},
  author = {Murad, Abdulmajid and Pyun, Jae-Young},
  year = {2017},
  month = nov,
  journal = {Sensors},
  volume = {17},
  number = {11},
  pages = {2556},
  doi = {10.3390/s17112556},
  urldate = {2024-09-03}
}

@misc{oordRepresentationLearningContrastive2019,
  title = {Representation {{Learning}} with {{Contrastive Predictive Coding}}},
  author = {van den Oord, Aaron and Li, Yazhe and Vinyals, Oriol},
  year = {2019},
  month = jan,
  number = {arXiv:1807.03748},
  eprint = {1807.03748},
  primaryclass = {cs, stat},
  publisher = {arXiv},
  doi = {10.48550/arXiv.1807.03748},
  urldate = {2024-09-14},
  archiveprefix = {arXiv}
}

@article{peiMicroexpressionRecognitionAlgorithm2019,
  title = {A {{Micro-expression Recognition Algorithm}} for {{Students}} in {{Classroom Learning Based}} on {{Convolutional Neural Network}}},
  author = {Pei, Jiayin and Shan, Peng},
  year = {2019},
  month = dec,
  journal = {Traitement du Signal},
  volume = {36},
  number = {6},
  pages = {557--563},
  doi = {10.18280/ts.360611},
  urldate = {2024-09-03}
}

@inproceedings{radfordLearningTransferableVisual2021a,
  title = {Learning {{Transferable Visual Models From Natural Language Supervision}}},
  booktitle = {Proceedings of the 38th {{International Conference}} on {{Machine Learning}}},
  author = {Radford, Alec and Kim, Jong Wook and Hallacy, Chris and Ramesh, Aditya and Goh, Gabriel and Agarwal, Sandhini and Sastry, Girish and Askell, Amanda and Mishkin, Pamela and Clark, Jack and Krueger, Gretchen and Sutskever, Ilya},
  year = {2021},
  month = jul,
  pages = {8748--8763},
  publisher = {PMLR},
  urldate = {2024-08-02}
}

@misc{robinsonContrastiveLearningHard2021,
  title = {Contrastive {{Learning}} with {{Hard Negative Samples}}},
  author = {Robinson, Joshua and Chuang, Ching-Yao and Sra, Suvrit and Jegelka, Stefanie},
  year = {2021},
  month = jan,
  number = {arXiv:2010.04592},
  eprint = {2010.04592},
  primaryclass = {cs, stat},
  publisher = {arXiv},
  doi = {10.48550/arXiv.2010.04592},
  urldate = {2024-09-14},
  archiveprefix = {arXiv}
}

@article{ronaoHumanActivityRecognition2016,
  title = {Human Activity Recognition with Smartphone Sensors Using Deep Learning Neural Networks},
  author = {Ronao, Charissa Ann and Cho, Sung-Bae},
  year = {2016},
  month = oct,
  journal = {Expert Systems with Applications},
  volume = {59},
  pages = {235--244},
  doi = {10.1016/j.eswa.2016.04.032},
  urldate = {2024-09-03}
}

@misc{sahooSystematicSurveyPrompt2024,
  title = {A {{Systematic Survey}} of {{Prompt Engineering}} in {{Large Language Models}}: {{Techniques}} and {{Applications}}},
  shorttitle = {A {{Systematic Survey}} of {{Prompt Engineering}} in {{Large Language Models}}},
  author = {Sahoo, Pranab and Singh, Ayush Kumar and Saha, Sriparna and Jain, Vinija and Mondal, Samrat and Chadha, Aman},
  year = {2024},
  month = feb,
  number = {arXiv:2402.07927},
  eprint = {2402.07927},
  primaryclass = {cs},
  publisher = {arXiv},
  doi = {10.48550/arXiv.2402.07927},
  urldate = {2024-09-07},
  archiveprefix = {arXiv}
}

@article{sikderKUHAROpenDataset2021,
  title = {{{KU-HAR}}: {{An}} Open Dataset for Heterogeneous Human Activity Recognition},
  shorttitle = {{{KU-HAR}}},
  author = {Sikder, Niloy and Nahid, Abdullah-Al},
  year = {2021},
  month = jun,
  journal = {Pattern Recognition Letters},
  volume = {146},
  pages = {46--54},
  doi = {10.1016/j.patrec.2021.02.024},
  urldate = {2024-09-07}
}

@inproceedings{sztylerOnbodyLocalizationWearable2016,
  title = {On-Body Localization of Wearable Devices: {{An}} Investigation of Position-Aware Activity Recognition},
  shorttitle = {On-Body Localization of Wearable Devices},
  booktitle = {2016 {{IEEE International Conference}} on {{Pervasive Computing}} and {{Communications}} ({{PerCom}})},
  author = {Sztyler, Timo and Stuckenschmidt, Heiner},
  year = {2016},
  month = mar,
  pages = {1--9},
  doi = {10.1109/PERCOM.2016.7456521},
  urldate = {2024-09-07}
}

@inproceedings{wangEffectiveYawnBehavior2019,
  title = {An {{Effective Yawn Behavior Detection Method}} in {{Classroom}}},
  booktitle = {Neural {{Information Processing}}},
  author = {Wang, Zexian and Jiang, Fei and Shen, Ruimin},
  editor = {Gedeon, Tom and Wong, Kok Wai and Lee, Minho},
  year = {2019},
  pages = {430--441},
  publisher = {Springer International Publishing},
  address = {Cham},
  doi = {10.1007/978-3-030-36708-4_35}
}

@article{xiaTS2ACTFewShotHuman2023a,
  title = {{{TS2ACT}}: {{Few-Shot Human Activity Sensing}} with {{Cross-Modal Co-Learning}}},
  shorttitle = {{{TS2ACT}}},
  author = {Xia, Kang and Li, Wenzhong and Gan, Shiwei and Lu, Sanglu},
  year = {2023},
  month = dec,
  journal = {Proceedings of the ACM on Interactive, Mobile, Wearable and Ubiquitous Technologies},
  volume = {7},
  number = {4},
  pages = {1--22},
  doi = {10.1145/3631445},
  urldate = {2024-09-15}
}

@inproceedings{yangStudentClassroomBehavior2023,
  title = {Student {{Classroom Behavior Detection Based}} on~{{YOLOv7}}+{{BRA}} and~{{Multi-model Fusion}}},
  booktitle = {Image and {{Graphics}}},
  author = {Yang, Fan and Wang, Tao and Wang, Xiaofei},
  editor = {Lu, Huchuan and Ouyang, Wanli and Huang, Hui and Lu, Jiwen and Liu, Risheng and Dong, Jing and Xu, Min},
  year = {2023},
  pages = {41--52},
  publisher = {Springer Nature Switzerland},
  address = {Cham},
  doi = {10.1007/978-3-031-46311-2_4}
}

@inproceedings{yuCCPoseNetHumanPose2023,
  title = {{{CC-PoseNet}}: {{Towards Human Pose Estimation}} in {{Crowded Classrooms}}},
  shorttitle = {{{CC-PoseNet}}},
  booktitle = {{{ICASSP}} 2023 - 2023 {{IEEE International Conference}} on {{Acoustics}}, {{Speech}} and {{Signal Processing}} ({{ICASSP}})},
  author = {Yu, Zefang and Hu, Yanping and Xiang, Suncheng and Liu, Ting and Fu, Yuzhuo},
  year = {2023},
  month = jun,
  pages = {1--5},
  doi = {10.1109/ICASSP49357.2023.10095734},
  urldate = {2024-09-03}
}

@article{yuPhysicalActivitySelfefficacy2024,
  title = {Physical Activity and Self-Efficacy in College Students: The Mediating Role of Grit and the Moderating Role of Gender},
  shorttitle = {Physical Activity and Self-Efficacy in College Students},
  author = {Yu, Hongyan and Zhu, Tingfei and Tian, Jianing and Zhang, Gang and Wang, Peng and Chen, Junxiong and Shen, Liqun},
  year = {2024},
  month = may,
  journal = {PeerJ},
  volume = {12},
  pages = {e17422},
  doi = {10.7717/peerj.17422},
  urldate = {2024-09-09}
}

@article{yuRawVideoPedagogical2024,
  title = {From {{Raw Video}} to {{Pedagogical Insights}}: {{A Unified Framework}} for {{Student Behavior Analysis}}},
  shorttitle = {From {{Raw Video}} to {{Pedagogical Insights}}},
  author = {Yu, Zefang and Xie, Mingye and Gao, Jingsheng and Liu, Ting and Fu, Yuzhuo},
  year = {2024},
  month = mar,
  journal = {Proceedings of the AAAI Conference on Artificial Intelligence},
  volume = {38},
  number = {21},
  pages = {23241--23249},
  doi = {10.1609/aaai.v38i21.30371},
  urldate = {2024-08-02}
}

@inproceedings{zbontarBarlowTwinsSelfSupervised2021a,
  title = {Barlow {{Twins}}: {{Self-Supervised Learning}} via {{Redundancy Reduction}}},
  shorttitle = {Barlow {{Twins}}},
  booktitle = {Proceedings of the 38th {{International Conference}} on {{Machine Learning}}},
  author = {Zbontar, Jure and Jing, Li and Misra, Ishan and LeCun, Yann and Deny, Stephane},
  year = {2021},
  month = jul,
  pages = {12310--12320},
  publisher = {PMLR},
  urldate = {2024-09-14}
}

@inproceedings{zengConvolutionalNeuralNetworks2014,
  title = {Convolutional {{Neural Networks}} for Human Activity Recognition Using Mobile Sensors},
  booktitle = {6th {{International Conference}} on {{Mobile Computing}}, {{Applications}} and {{Services}}},
  author = {Zeng, Ming and Nguyen, Le T. and Yu, Bo and Mengshoel, Ole J. and Zhu, Jiang and Wu, Pang and Zhang, Joy},
  year = {2014},
  month = nov,
  pages = {197--205},
  doi = {10.4108/icst.mobicase.2014.257786},
  urldate = {2024-09-03}
}

@inproceedings{zhengIntelligentStudentBehavior2020,
  title = {Intelligent {{Student Behavior Analysis System}} for {{Real Classrooms}}},
  booktitle = {{{ICASSP}} 2020 - 2020 {{IEEE International Conference}} on {{Acoustics}}, {{Speech}} and {{Signal Processing}} ({{ICASSP}})},
  author = {Zheng, Rui and Jiang, Fei and Shen, Ruimin},
  year = {2020},
  month = may,
  pages = {9244--9248},
  doi = {10.1109/ICASSP40776.2020.9053457},
  urldate = {2024-08-26}
}

@inproceedings{zhouWhoAreRaising2018,
  title = {Who {{Are Raising Their Hands}}? {{Hand-Raiser Seeking Based}} on {{Object Detection}} and {{Pose Estimation}}},
  shorttitle = {Who {{Are Raising Their Hands}}?},
  booktitle = {Proceedings of {{The}} 10th {{Asian Conference}} on {{Machine Learning}}},
  author = {Zhou, Huayi and Jiang, Fei and Shen, Ruimin},
  year = {2018},
  month = nov,
  pages = {470--485},
  publisher = {PMLR},
  urldate = {2024-09-03}
}

@inproceedings{li-etal-2025-sensorllm,
    title = "{S}ensor{LLM}: Aligning Large Language Models with Motion Sensors for Human Activity Recognition",
    author = "Li, Zechen  and
      Deldari, Shohreh  and
      Chen, Linyao  and
      Xue, Hao  and
      Salim, Flora D.",
    booktitle = "Proceedings of the 2025 Conference on Empirical Methods in Natural Language Processing",
    year = "2025",
    publisher = "Association for Computational Linguistics",
    url = "https://aclanthology.org/2025.emnlp-main.19/",
    pages = "354--379",
}

@article{colakhodzicIdentifyingHomogenousGroups2012,
  title = {Identifying Homogenous Groups Regarding Situational-Motor Abilities of Young Football Players},
  author = {Colakhodzic, Ekrem and Radjo, Izet and Talovi{\'c}, Munir and La{\v c}i{\'c}, Osman},
  year = 2012,
  month = jun,
  journal = {Homosporticus},
  volume = {Volume 14},
  pages = {pg.33-40.}
}

@article{dysonUsingCooperativeLearning2001,
  title = {Using {{Cooperative Learning Structures}} in {{Physical Education}}},
  author = {Dyson, Ben and Grineski, Steve},
  year = 2001,
  month = feb,
  journal = {Journal of Physical Education, Recreation \& Dance},
  publisher = {Taylor \& Francis Group},
  issn = {0730-3084},
  url = {https://www.tandfonline.com/doi/abs/10.1080/07303084.2001.10605831},
  urldate = {2026-01-19},
  copyright = {Copyright Taylor and Francis Group, LLC},
  langid = {english}
}

@article{weiTestingEffectsPlayer2025,
  title = {Testing {{The Effects}} of {{Player Matching}} in {{Basketball Matches}} and {{Small-Sided Game Training Scenarios Based}} on {{The Relative Age Effect}}: {{A}} 6-{{Month Study}} on {{Physical Performance}} and {{Skill Adaptations}}},
  shorttitle = {Testing {{The Effects}} of {{Player Matching}} in {{Basketball Matches}} and {{Small-Sided Game Training Scenarios Based}} on {{The Relative Age Effect}}},
  author = {Wei, LiXin and Zheng, Yafei and Li, MingBang and Dai, Shu},
  year = 2025,
  month = jun,
  journal = {Journal of Sports Science and Medicine},
  volume = {24},
  pages = {397--405},
  doi = {10.52082/jssm.2025.397}
}
